\documentclass[fleqn,usenatbib]{mnras}
\usepackage{mathptmx}
\usepackage[varvw]{newtxmath}  
\usepackage[T1]{fontenc}
\usepackage{ae,aecompl}
\usepackage{times}
\usepackage{amsmath}
\usepackage{upgreek}
\usepackage{xcolor}
\usepackage{graphicx}
\usepackage{pdflscape}
\usepackage{multicol} 
\usepackage{amsmath,bm}
\usepackage{caption}

\begin{document}

\title{Molecular clouds as hubs in spiral galaxies : gas inflow and evolutionary sequence}
\author[J. W. Zhou]{
J. W. Zhou \thanks{E-mail: jwzhou@mpifr-bonn.mpg.de}$^{1}$
Sami Dib $^{2}$
Timothy A. Davis $^{3}$
\\
$^{1}$Max-Planck-Institut f\"{u}r Radioastronomie, Auf dem H\"{u}gel 69, 53121 Bonn, Germany\\
$^{2}$Max Planck Institute f\"{u}r Astronomie, K\"{o}nigstuhl 17, 69117 Heidelberg, Germany\\
$^{3}$Cardiff Hub for Astrophysics Research \& Technology, School of Physics \& Astronomy, Cardiff University, Queens Buildings, Cardiff CF24 3AA, UK
}

\date{Accepted XXX. Received YYY; in original form ZZZ}

\pubyear{2024}
\maketitle

\begin{abstract}
We decomposed the molecular gas in the spiral galaxy NGC 628 (M74) into multi-scale hub-filament structures using the CO (2$-$1) line by the dendrogram algorithm. All leaf structures as potential hubs were classified into three categories, i.e. leaf-HFs-A, leaf-HFs-B and leaf-HFs-C. leaf-HFs-A exhibit the best hub-filament morphology, which also have the highest density contrast, the largest mass and the lowest virial ratio. We employed the FILFINDER algorithm to identify and characterize filaments within 185 leaf-HFs-A structures, and fitted the velocity gradients around the intensity peaks. Measurements of velocity gradients provide evidence for gas inflow within these structures, which can serve as a kinematic evidence that these structures are hub-filament structures.
The numbers of the associated 21 $\mu$m and H$_{\alpha}$ structures and the peak intensities of 7.7 $\mu$m, 21 $\mu$m and H$_{\alpha}$ emissions decrease from leaf-HFs-A to leaf-HFs-C. The spatial separations between the intensity peaks of CO and 21 $\mu$m structures of leaf-HFs-A are larger than those of leaf-HFs-C. These evidence indicate that leaf-HFs-A are more evolved than leaf-HFs-C.
There may be an evolutionary sequence from leaf-HFs-C to leaf-HFs-A. Currently, leaf-HFs-C lack a distinct gravitational collapse process that would result in a significant density contrast. The density contrast can effectively measure the extent of the gravitational collapse and the depth of the gravitational potential of the structure which, in turn, shapes the hub-filament morphology. Combined with the kinematic analysis presented in previous studies, 
a picture emerges that molecular gas in spiral galaxies is organized into network structures through the gravitational coupling of multi-scale hub-filament structures. Molecular clouds, acting as knots within these networks, serve as hubs, which are local gravitational centers and the main sites of star formation.
\end{abstract}


\maketitle 

\section{Introduction} 

Analyzing the dynamical interaction between density enhancements in giant molecular clouds and gas motion in their surrounding environment provides insight into the formation of hierarchical structures in high-mass star and star cluster- forming regions
\citep{McKee2007-45, Dib2012-758, Motte2018-56, Henshaw2020-4}. 
Detailed observations of high-mass star-forming regions with high resolution unveil the organized distribution of density enhancements within filamentary networks of gas, particularly evident in hub-filament systems. In these systems, converging flows channel material towards the central hub along the interconnected filaments
\citep{Peretto2013,Henshaw2014,Zhang2015,Liu2016,Yuan2018,Lu2018,Issac2019,Dewangan2020,Liu2022-511,Zhou2022-514,Dib2023-524, Zhou2023-676}.
In particular, \citet{Zhou2022-514} studied the physical properties and evolution of hub-filament systems across $\sim$ 140 protoclusters using spectral line data obtained from the ATOMS
(ALMA Three-millimeter Observations of Massive Star-forming regions) survey \citep{Liu2020}.
They proposed that hub-filament structures exhibiting self-similarity and filamentary accretion appear to persist across a range of scales within high-mass star-forming regions, spanning from several thousand astronomical units to several parsecs. This paradigm of hierarchical, multi-scale hub-filament structures was generalized from clump-core scale to cloud-clump scale in \citet{Zhou2023-676} and from cloud-clump scale to galaxy-cloud scale in \citet{Zhou2024arXiv}. 
Hierarchical collapse and hub-filament structures feeding the central regions are also  depicted in previous works, 
see \citet{Motte2018-56,Vazquez2019-490, Kumar2020-642} and references therein.

Kinematic analyses presented in \citet{Zhou2023-676} and \citet{Zhou2024arXiv} demonstrate the presence of multi-scale hub-filament structures within molecular clouds and spiral galaxies. The results notably show that intensity peaks, acting as hubs, are correlated with converging velocities, suggesting that surrounding gas flows are directed towards these dense regions. Filaments across various scales exhibit distinct velocity gradients, with a marked increase in these gradients at smaller scales. 
Interestingly, the variations in velocity gradients measured at larger scales align with expectations from gravitational free-fall with higher central masses.
This correlation implies that inflows on large scales are driven by large-scale structures, potentially due to the gravitational coupling of smaller-scale structures. 
Fig.~5 of \citet{Zhou2024arXiv} shows a vivid example of gravitational coupling, where multiple peaks are coupled together to form a gravitational potential well on larger scale, and each peak itself is also a local gravitational center. This aligns with the global hierarchical collapse (GHC) scenario proposed by \citet{Vazquez2019-490}, which suggests that clouds are composed of multiple nested collapses occurring across a wide range of scales.
These observations agree with the hierarchical nature found in molecular clouds and spiral galaxies, and gas inflow from large to small scales. 
Large-scale velocity gradients are consistently associated with numerous intensity peaks, reinforcing the idea that the clustering of smaller-scale structures can act as gravitational centers on larger scales.

Based on the kinematic evidence, the main goal of this work is to directly recover the multi-scale hub-filament structures in a spiral galaxy NGC 628 (M74 or the Phantom Galaxy). 

\section{Data} 

We selected a face-on spiral galaxy NGC 628 (M74) from the PHANGS-ALMA survey. We used the combined 12m+7m+TP PHANGS-ALMA CO (2$-$1) data cubes to investigate gas kinematics and dynamics and identify hub-filament structures in the galaxy, which have a spectral resolution of 2.5 km s$^{-1}$ and an angular resolution $\sim$1.1$''$, corresponding
to a linear resolution $\sim$50 pc at the distance 9.8 Mpc \citep{Leroy2021-257,Anand2021-501}.
We also use the James Webb Space Telescope (JWST) 7.7 and 21 $\mu$m maps with an angular resolution $\sim$0.67$''$ or a linear resolution $\sim$30 pc from the PHANGS-JWST survey and the H$_{\alpha}$ emission map with an angular resolution $\sim$0.92$''$ or a linear resolution $\sim$45 pc from the PHANGS-MUSE survey.
Overviews of the PHANGS-ALMA, PHANGS-MUSE and PHANGS-JWST  surveys' science goals, sample selection, observation strategy, and data products are described in \citet{Leroy2021-257,Leroy2021-255,Emsellem2022-659,Lee2023-944}. 
All the data are available on the PHANGS team website \footnote{\url{https://sites.google.com/view/phangs/home}}. 
The field-of-views (FOVs) of these observations are shown in Fig.~\ref{co}. 
NGC 628 was selected for the following reasons:

1. It is a face-on galaxy. The inclination angle of NGC 628 is only 8.9 degree \citep{Lang2020-897}. Therefore, this facilitates the identification of the hub-filament structures embedded within the galaxy.

2. High-resolution CO data and the galaxy's relatively near distance can clearly reveal the hub-filament structures.

3. It is covered by multi-wavelength observations. As discussed below, high-resolution 21 $\mu$m and H$_{\alpha}$ emission are crucial to determine the evolutionary states of CO structures.

\section{Results} 

\subsection{Dendrogram}\label{dendro}

\begin{figure}
\centering
\includegraphics[width=0.5\textwidth]{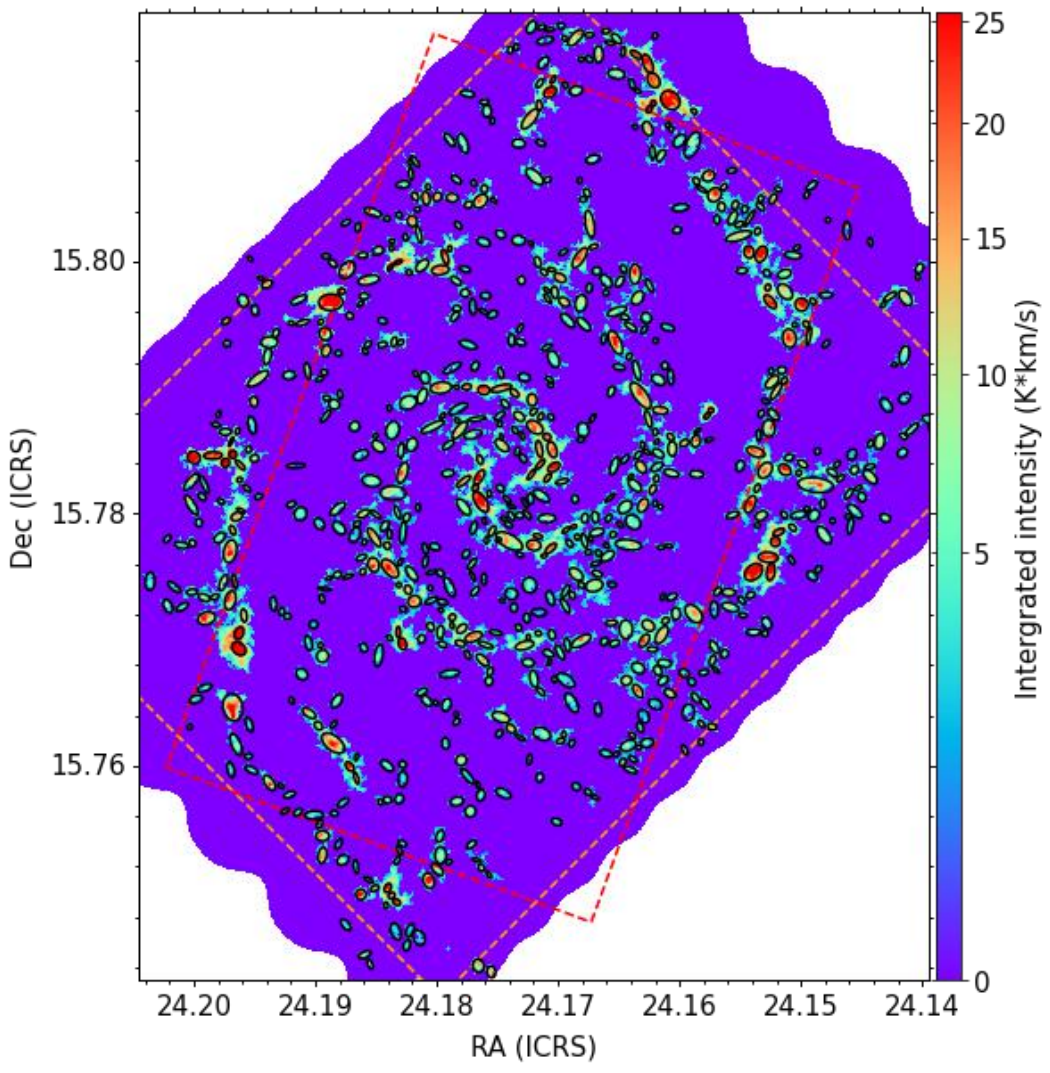}
\caption{Background is the intergrated intensity map of CO (2$-$1) emission. Black ellipses represent leaf structures identified by the dendrogram algorithm. Red and orange boxes mark the field-of-views in 21 $\mu$m and H$_{\alpha}$ observations, respectively.
The long and short axes of the ellipse are $2a$ and $2b$, as discussed in Sec.\ref{dendro}.}
\label{co}
\end{figure}
\begin{figure}
\centering
\includegraphics[width=0.5\textwidth]{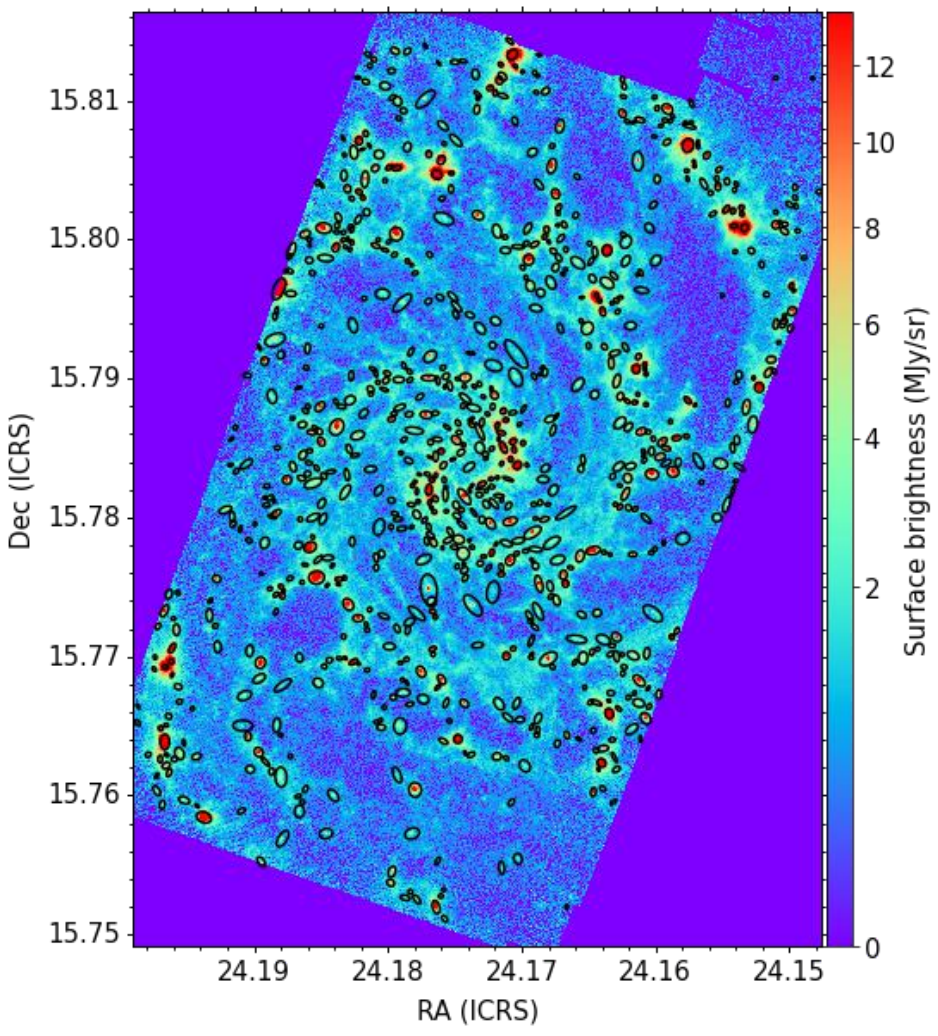}
\caption{Background is the intensity map of JWST 21 $\mu$m emission. Black ellipses represent the remained (bright) 21 $\mu$m structures. The selection was done in Sec.\ref{association}. The long and short axes of the ellipse are $2a$ and $2b$.}
\label{21}
\end{figure}
\begin{figure}
\centering
\includegraphics[width=0.5\textwidth]{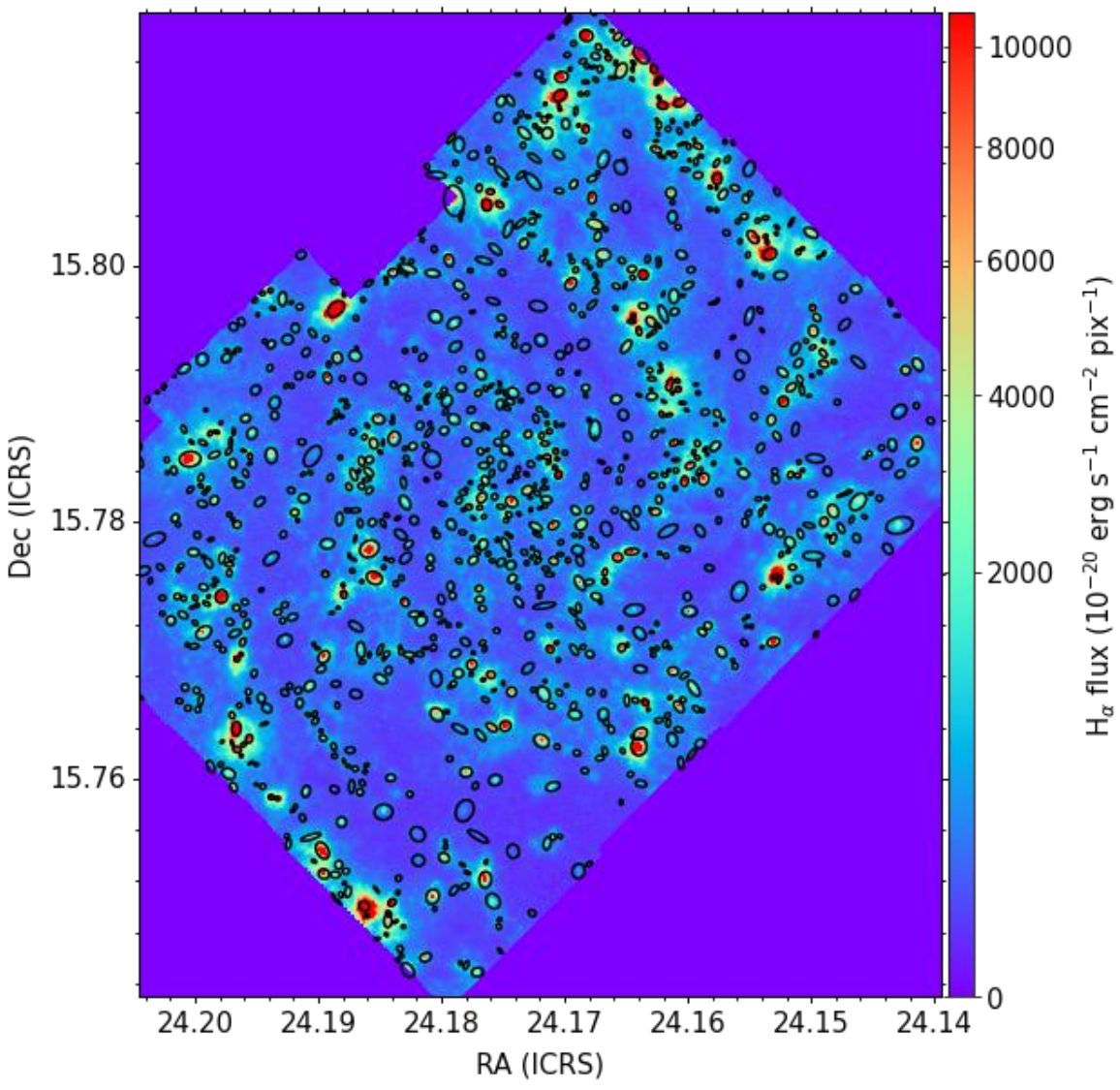}
\caption{Background is the intensity map of H$_{\alpha}$ emission. Black ellipses represent the remained (bright) H$_{\alpha}$ structures. The selection was done in Sec.\ref{association}. The long and short axes of the ellipse are $2a$ and $2b$.}
\label{ha}
\end{figure}

We conducted a direct identification of hierarchical (sub-)structures based on the 2D intensity maps. As described in \citet{Rosolowsky2008-679}, the dendrogram algorithm decomposes density or intensity data into hierarchical structures called leaves, branches, and trunks. 
Using the {\it astrodendro} package \footnote{\url{https://dendrograms.readthedocs.io/en/stable/index.html}},
there are three major input parameters for the dendrogram algorithm: {\it min\_value} for the minimum value to be considered in the dataset, {\it min\_delta} for a leaf that can be considered as an independent entity, and {\it min\_npix} for the minimum area of a structure.
For the CO (2$-$1) data cube, 
there are two types of Moment 0 maps (strictly masked and broadly masked) in the data product of the PHANGS-ALMA survey \footnote{Details of the masking strategy and completeness statistics are presented in the PHANGS pipeline paper \citep{Leroy2021-255}.}. The strictly masked maps only include emissions that are identified as signals with high confidence in the data cube, which might filter out the relatively faint structures. The broadly masked maps offer superior completeness and cover larger areas compared to the strictly masked maps. However, due to the inclusion of more regions with faint emissions or areas close to bright emissions, they tend to be noisier and may contain false positives. 
In order to ensure the reliability of the identified structures and because we are only interested in local dense structures, we select the strictly masked Moment 0 map to identify structures.
Since all the retained structures on the strictly masked Moment 0 map are reliable, we only require the smallest area of the identified structure be larger than 1 beam area. We do not set additional parameters in the algorithm to minimize the dependence of the identification on parameter settings. Finally, we obtained 773 leaf structures.

For the 21 $\mu$m and H$_{\alpha}$ emission, apart from {\it min\_npix}= 1 beam area, we also take 
the values of {\it min\_value}= 3*$\sigma_{\rm rms}$, {\it min\_delta} = 3*$\sigma_{\rm rms}$, where $\sigma_{\rm rms}$ is the background intensity. The total numbers of 21 $\mu$m and H$_{\alpha}$ leaf structures are 1491 and 1965, respectively.
We first retained as much structures as possible using these lower standards, then eliminated the diffuse structures, as described in Sec.~\ref{association}.
In Fig.~\ref{co}, Fig.~\ref{21} and Fig.~\ref{ha}, 
the CO, 21 $\mu$m and H$_{\alpha}$ structures identified by the dendrogram algorithm exhibit a strong correspondence with the background intensity maps.

The algorithm characterizes the morphology of each structure by approximating it as an ellipse. Within the dendrogram, the root mean square (rms) sizes (second moments) of the intensity distribution along the two spatial dimensions define the long and short axes of the ellipse, denoted as $a$ and $b$. As described in \citet{Zhou2024-682}, the initial ellipse with $a$ and $b$ is smaller, so a multiplication factor of two is applied to appropriately enlarge the ellipse.
Then the effective physical radius of an ellipse is $R_{\rm eff}$ =$\sqrt{2a \times 2b}*D$, where $D$ is the distance of the galaxy.
For a structure with an area $A$ and a total integrated intensity $I_{\rm CO}$, the mass of the structure can be calculated by
\begin{equation}
M = \alpha^{2-1}_{\rm CO} \times I_{\rm CO} \times A,
\label{mass}
\end{equation}
where $\alpha^{2-1}_{\rm CO} \approx 6.7~\rm M_{\odot} \rm pc^{-2} (\rm K~km~s^{-1})^{-1}$ \citep{Leroy2022-927}.

\subsection{Velocity components}\label{component}

\begin{figure*}
\centering
\includegraphics[width=1\textwidth]{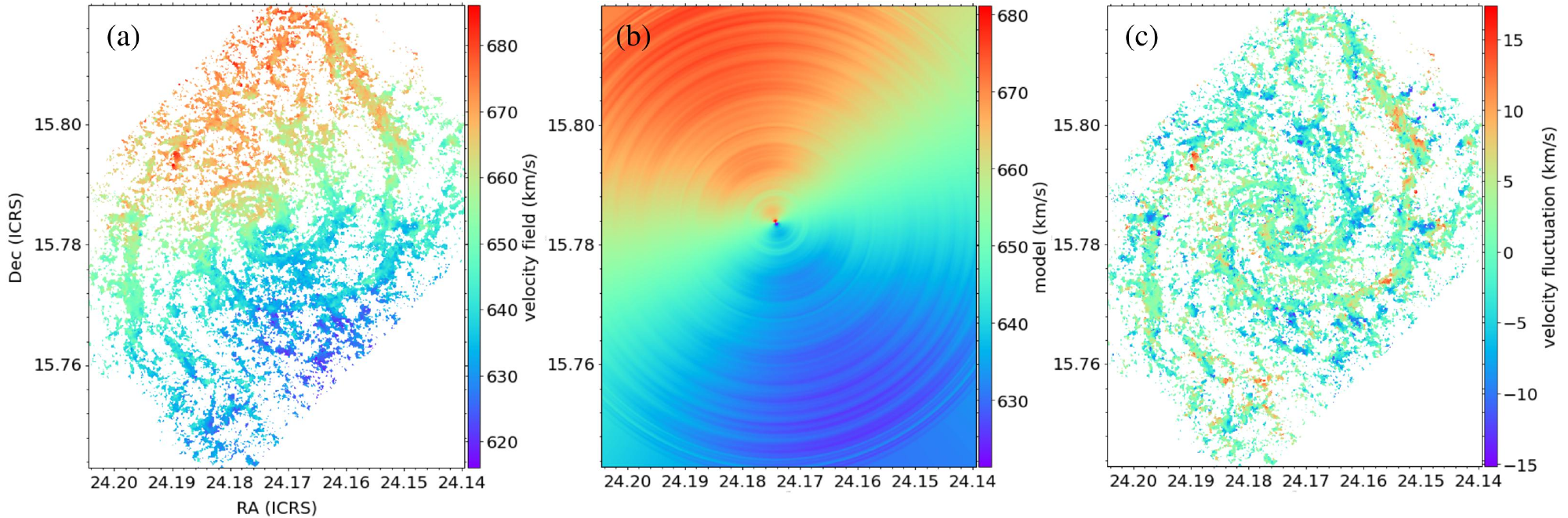}
\caption{(a) The velocity field (Moment 1) of NGC 628 in the observation; (b) Gas dynamical model created by the Kinematic Molecular Simulation (KinMS) package; (c) Velocity fluctuations after subtracting the model.}
\label{model}
\end{figure*}

\begin{figure*}
\centering
\includegraphics[width=1\textwidth]{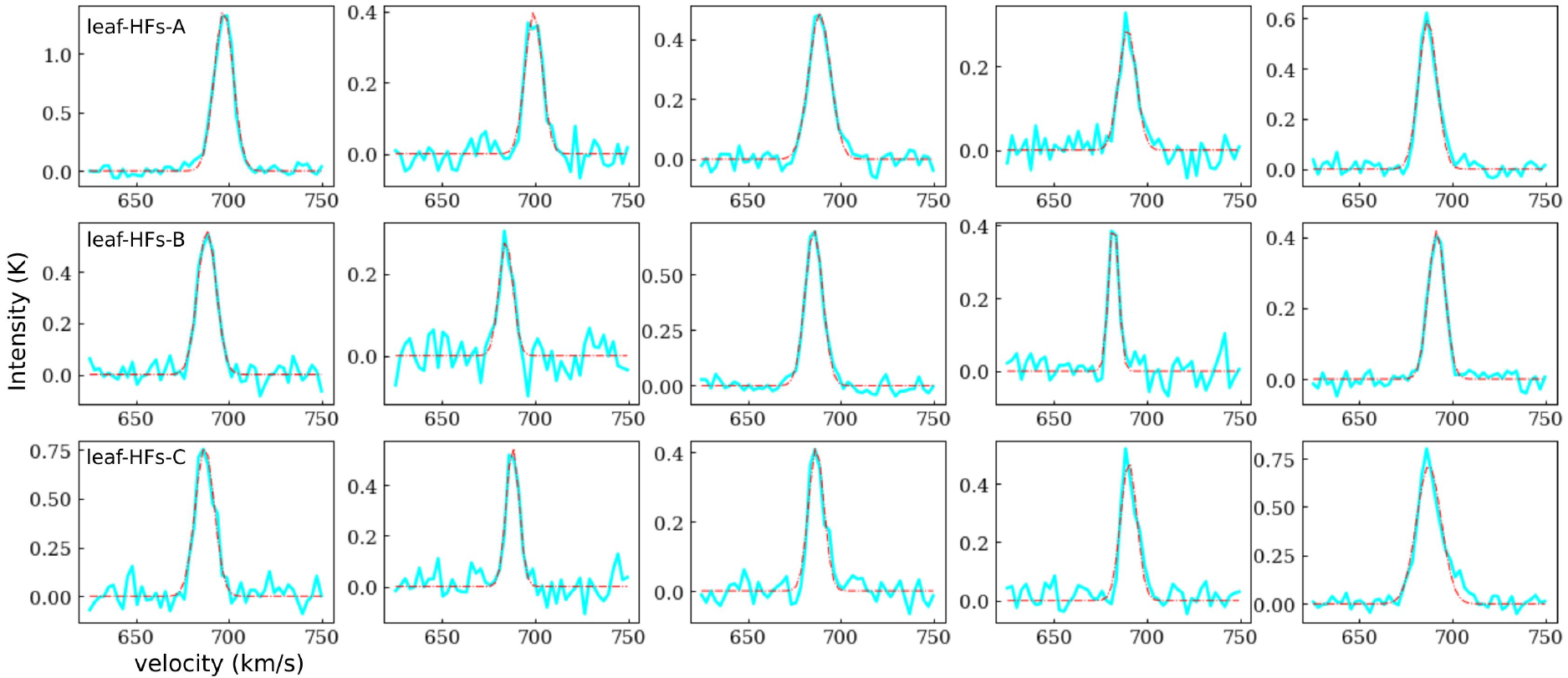}
\caption{
Average spectra and their Gaussian fitting of the CO structures marked by cyan ellipses in Fig.~\ref{case-A} (the first row), Fig.\ref{case-B} (the second row) and Fig.~\ref{case-C} (the third row). 
}
\label{spectra}
\end{figure*}
For the identified molecular structures, we extracted the average spectrum of each structure to investigate its velocity components and gas kinematics. Large-scale velocity gradients from the galaxy's rotation contribute to the non-thermal velocity dispersion. To address this, and before extracting the average spectra, we removed the bulk motion due to the rotation of galaxy by creating gas dynamical models using the Kinematic Molecular Simulation (KinMS) package \citep{Davis2013-429}, as shown in Fig.~\ref{model}.  
Then, following the procedure described in \citet{Zhou2024-682}, we fitted the averaged spectra of 773 leaf structures individually using the fully automated Gaussian decomposer \texttt{GAUSSPY+} \citep{Lindner2015-149, Riener2019-628} algorithm. 
Almost all structures exhibit a single-peak profile. Only a few structures have a clear double-peak profile. Fig.~\ref{spectra} displays the average spectra of the central regions of the structures presented in Fig.~\ref{case-A}, Fig.~\ref{case-B} and Fig.~\ref{case-C}, marked by cyan ellipses.

\subsection{Classification}\label{class}

\begin{figure*}
\centering
\includegraphics[width=1\textwidth]{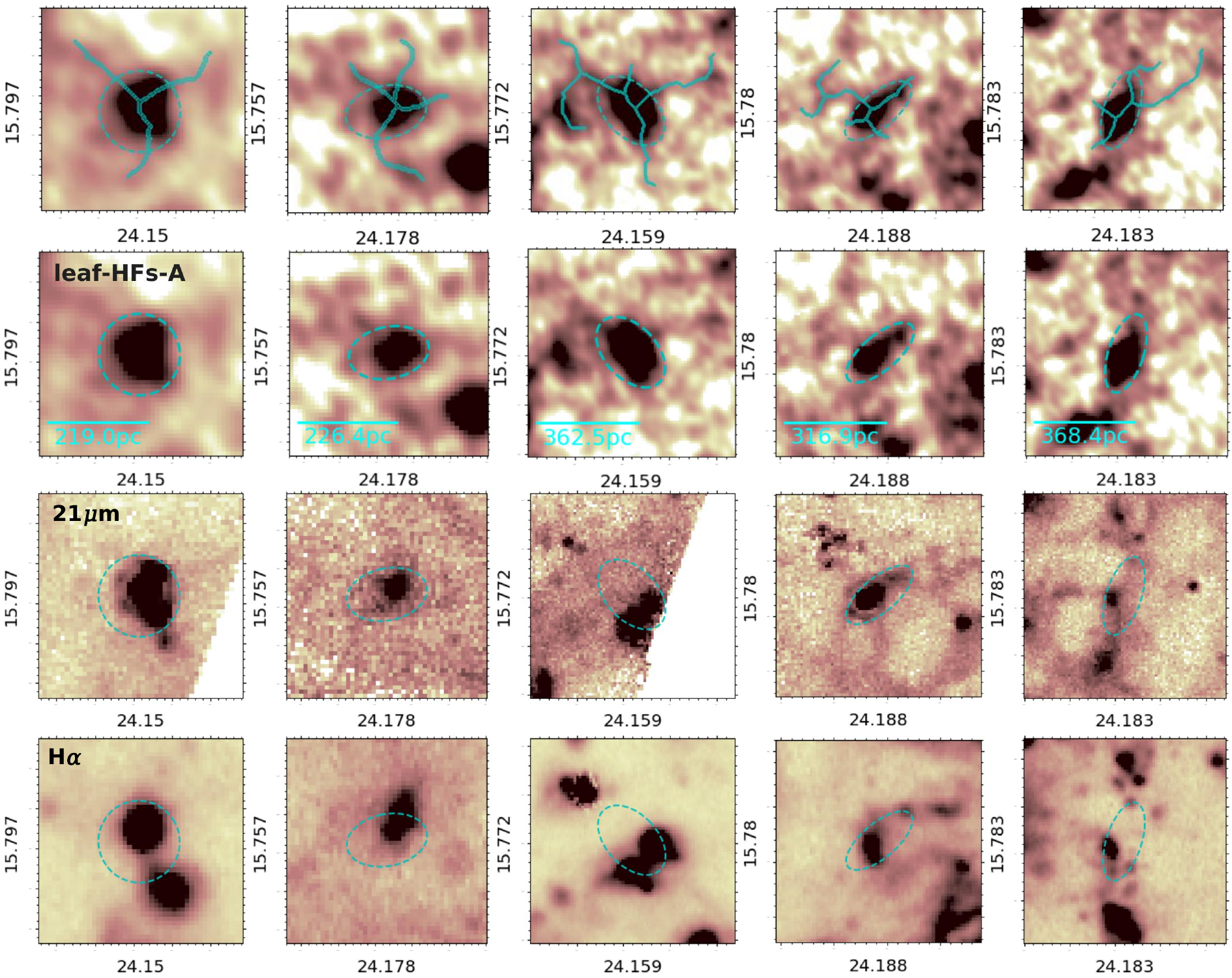}
\caption{Some examples of leaf-HFs-A. CO emission
(the first and second rows), 21 $\mu$m emission (the second row) and H$_{\alpha}$ emission (the third row).
The long and short axes of the ellipse are $2a$ and $2b$, as discussed in Sec.\ref{dendro}.
In the first row, the filaments identified by the FILFINDER algorithm are overlaid on the Moment 0 maps.}
\label{case-A}
\end{figure*}

\begin{figure*}
\centering
\includegraphics[width=1\textwidth]{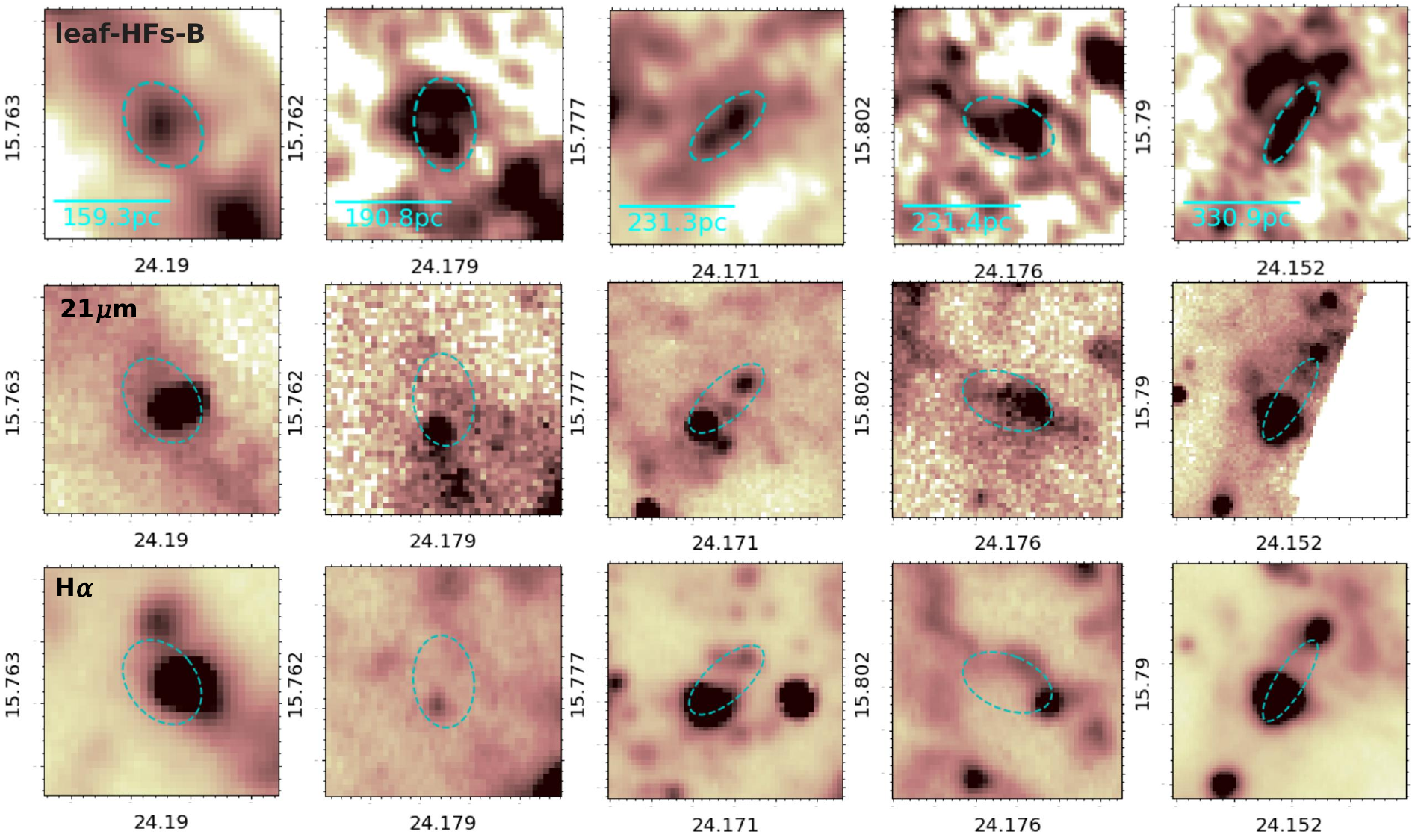}
\caption{Same as Fig.\ref{case-A}, but for leaf-HFs-B.}
\label{case-B}
\end{figure*}

\begin{figure*}
\centering
\includegraphics[width=1\textwidth]{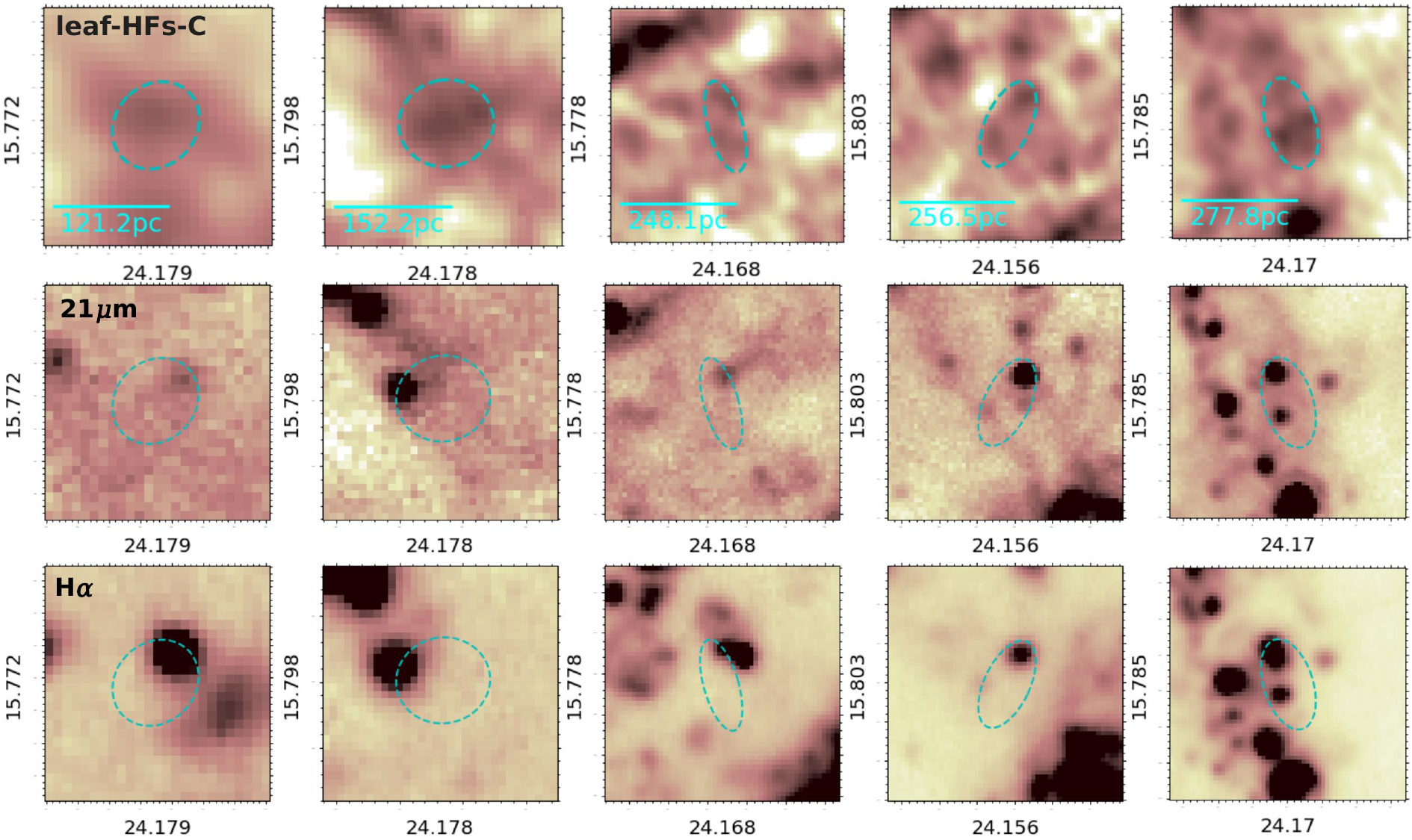}
\caption{Same as Fig.\ref{case-A}, but for leaf-HFs-C.}
\label{case-C}
\end{figure*}

From the line profile, we can fix the velocity range of each structure. Then, the Moment 0 map of each structure was reproduced in the corresponding velocity range to eliminate the overlap of potential incoherent velocity components. 
All identified intensity peaks (leaf structures) of CO (2$-$1) emission are thought to be potential hubs. 
Around these intensity peaks, we extended the spatial ranges to investigate the filamentary structures connected with them. After trying different enlargement factors, extending to 2.5 times the hub size (the effective radius of the leaf structure), we can recover the entire hub-filament structure and at the same time, avoid including many neighboring structures, as shown in Fig.~\ref{case-A}, Fig.~\ref{case-B} and Fig.~\ref{case-C}. 

All leaf structures were classified by eye into three categories only based on their morphology, i.e. leaf-HFs-A, leaf-HFs-B and leaf-HFs-C. The numbers of structures in the three categories are 234, 181 and 358. Some examples in the three categories are shown in Fig.~\ref{case-A}, Fig.~\ref{case-B} and Fig.~\ref{case-C}.
From these maps, we can observe that leaf-HFs-C do not exhibit clear central hubs, meaning that the density contrast between the hub and the surrounding diffuse gas is not pronounced. In contrast, leaf-HFs-A and leaf-HFs-B feature distinct central hubs or high-density central regions, characteristic of hub-filament structures. Compared to leaf-HFs-B, leaf-HFs-A demonstrate more defined hub-filament morphology, with filamentary diffuse gas surrounding their central hubs.
The hubs of leaf-HFs-A are more outstanding.
While the boundary between leaf-HFs-A and leaf-HFs-B might be less distinct, the difference between leaf-HFs-A and leaf-HFs-C is significant.
Therefore, in the subsequent discussion, we focus on the comparison between leaf-HFs-A and leaf-HFs-C.

The initial classification was only based on the morphology, because other differences between the structures were unknown. In the subsequent analysis,
we can see that the physical properties of the structures in three categories are also significantly different. Therefore, the morphological differences are the result of certain physical processes shaping them and are not coincidental.

\subsection{Filamentary structures}\label{filament}

\subsubsection{Identification}

As done in \citet{Zhou2024arXiv}, we also use the FILFINDER algorithm to characterize the filamentary structures around the hub. 
In \citet{Zhou2024arXiv}, we only focused on large-scale filaments along the spiral arms. Clear fluctuations in velocity and density were observed along these filaments. Each individual intensity peak reveals a local hub. In this work, we first identified the local intensity peaks (leaf structures), and then searched for hub-filament structures centered around these intensity peaks as potential hubs. 
Now, we need to continue identifying small-scale filamentary structures within these local hub-filament structures. Due to the limited observational resolution,
even for the largest leaf-HFs-A structures, they still lack enough pixels to recognize the filaments. Therefore, we have to regrid the images and increase the number of pixels in the images from $N_{x}*N_{y}$ to $2N_{x}*2N_{y}$ \footnote{We do this by using the \textbf{zoom} function from the \textbf{scipy.ndimage} module to regrid the image by setting an interpolation order of 3.}. To avoid introducing false structures, we only doubled the number of pixels. As shown in Fig.~\ref{case-A}, the morphology of the structures remains unchanged. The identified filamentary structures also align well with the background. However, due to the resolution limitations, the filamentary structures do not appear very extended.

\subsubsection{Velocity gradient}

\begin{figure}
\centering
\includegraphics[width=0.46\textwidth]{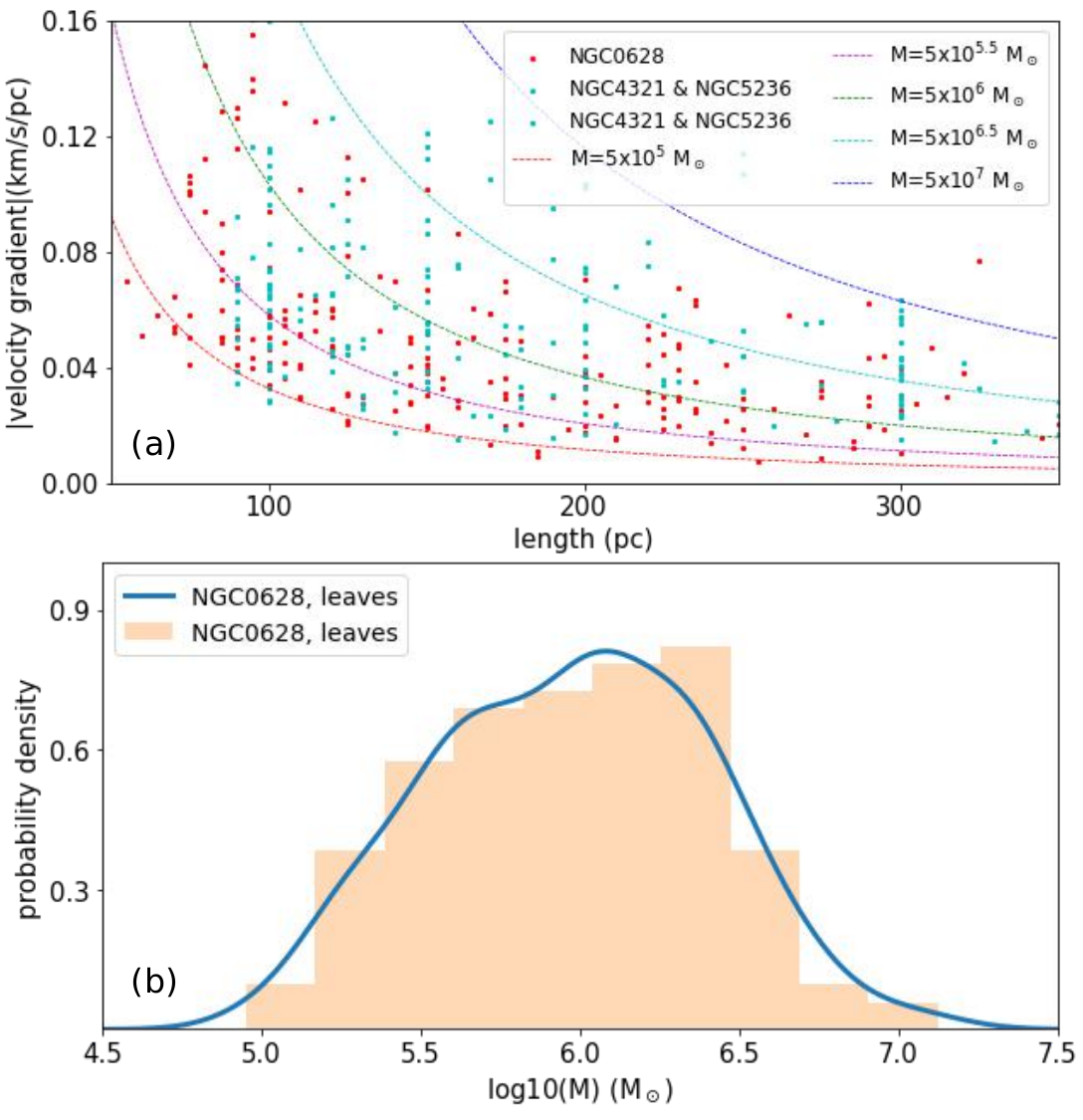}
\caption{
Statistical analysis of all fitted velocity gradients.
(a) Velocity gradient vs. the length. The color lines show the freefall velocity gradients for comparison. For the freefall model, red, magenta, green, cyan, and blue lines denote masses of 10$^{5}$\,M$_\odot$, 10$^{5.5}$\,M$_\odot$, 10$^{6}$\,M$_\odot$ , 10$^{6.5}$\,M$_\odot$ and 10$^{7}$\,M$_\odot$, respectively;
(b) The mass distribution of leaf structures.}
\label{gradient}
\end{figure}

A kinematic feature of hub-filament structures is that the gas flow along the filament converges towards the central hub, resulting in a measurable velocity gradient along the filament. Since we are now studying each local hub-filament structure, we follow the same analysis presentend in \citet{Zhou2022-514} to fit the velocity gradients around the intensity peaks. In Fig.~\ref{gradient}(a), the velocity gradients fitted in NGC 628 are quite comparable to those in NGC 4321 and NGC 5236 presented in \citet{Zhou2024arXiv}. 
The measured velocity gradients predominantly fit within the mass range of approximately $10^{5}$ to $10^{7}$ M$_\odot$. This range aligns with the mass distribution of leaf structures shown in Fig.~\ref{gradient}(b). This consistency suggests that local dense structures act as gravitational centers, accreting surrounding diffuse gas and thereby generating the observed velocity gradients. 
Measurements of velocity gradients provide evidence for gas inflow within these structures, which can also serve as kinematic evidence that these structures can be regarded as hub-filament systems.

We note that only leaf-HFs-A structures were considered in the identification of filaments and the analysis of the velocity gradients. Furthermore, we successfully identified the filaments for only 185 leaf-HFs-A structures, and the filaments are long enough to show clear velocity gradients.

\subsection{Physical properties}

\subsubsection{Density contrast}

\begin{figure*}
\centering
\includegraphics[width=1\textwidth]{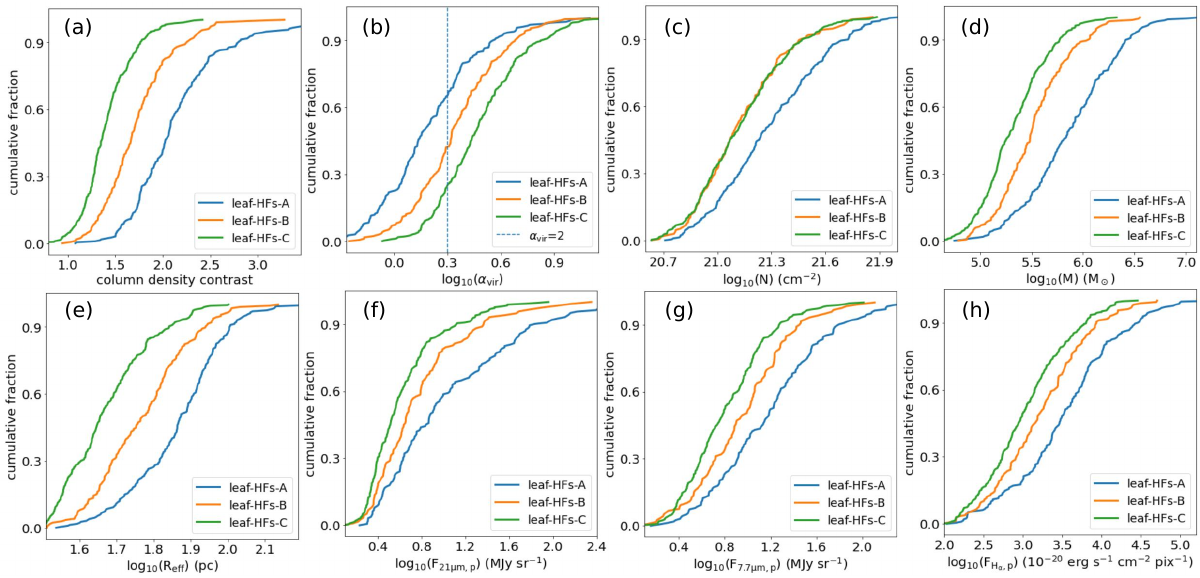}
\caption{Physical parameters of three types of structures. (a) Column density contrast; (b) Virial ratio. The vertical dashed line marks the position of $\alpha_{vir}=2$; (c) Mass; (d) Column density; (e) Scale (the effective radius of the CO structures); (f) Peak intensity of 7.7 $\mu$m emission; (g) Peak intensity of 21 $\mu$m emission; (h) Peak intensity of H$_{\alpha}$ emission.}
\label{cp0}
\end{figure*}

As shown in Fig.\ref{case-A}, a clear central hub implies a noticeable density contrast between the hub and the surrounding diffuse gas. We define the density contrast $C$ as the ratio of the average column density in the hub region, $N_{\rm hub}$, to the average column density within an elliptical ring around the hub with the width equal to the hub size (the effective radius of the leaf structure), $N_{\rm a}$, 
\begin{equation}
C=N_{\rm hub}/N_{\rm a}. 
\end{equation}
As expected, leaf-HFs-A with the best hub-filament morphology have the highest density contrast in Fig.~\ref{cp0}(a). Moreover, leaf-HFs-A also possess the largest masses and the lowest virial ratios.

\subsubsection{Association with 21 $\mu$m and H$_{\alpha}$ emission}\label{association}


Diffuse 21 $\mu$m and H$_{\alpha}$ emission should be unrelated to recent star formation activities. Therefore, for 21 $\mu$m and H$_{\alpha}$ structures identified in Sec.~\ref{dendro}, we need to filter the faint structures.
21 $\mu$m and H$_{\alpha}$ emission that originate from the clusters born in the clouds should present intensity peaks. Similar to the CO structures, we calculated the intensity contrast for all identified 21 $\mu$m and H$_{\alpha}$ structures.
Before the calculation, we first shift the center of each structure to the pixel with the strongest emission. 
Structures with peak intensities greater than the median peak intensity of all structures are retained. 
Given that a large number of faint structures were identified using loose criteria in Sec.\ref{dendro}, this standard is not stringent.
For structures that do not meet this criterion, if their intensity contrast is greater than 1.5, they will also be retained. 
The final numbers of 21 $\mu$m and H$_{\alpha}$ leaf structures are 841 and 1142, respectively.
As can be seen from Fig.~\ref{21} and Fig.~\ref{ha}, the retained (bright) structures encompass all the significant emissions.

As shown in Fig~.\ref{co}, JWST 21 $\mu$m emission has the smallest field-of-view in the observations.
Thus, we discarded the CO (2$-$1) and H$_{\alpha}$ structures which are beyond the FOV of the 21 $\mu$m observation.
Generally, the structure seen in CO is irregular and extended, so the densest part may not be at the effective center of the structure output by the dendrogram algorithm. However, the position with the highest column density truly represents the site of star formation. Therefore, 
we calculate the spatial separations between the position of maximum CO column density and the positions of maximum 21 $\mu$m or H$_{\alpha}$ intensity.
If the separations are less than the effective radius of a CO structure, we consider the corresponding 21 $\mu$m or H$_{\alpha}$ structures are associated with the CO structure. 
We note that CO could be optically thick, which may lead to an underestimation of the column density and, consequently, affect the estimation of the density center.

Finally, there are 313 21 $\mu$m structures (37\%, 313/841) and 324 H$_{\alpha}$ structures (28\%, 324/1142) associated with CO structures, where 185 CO structures have both strong 21 $\mu$m and H$_{\alpha}$ emissions. 
Limited to the FOV of 21 $\mu$m observations, the total numbers of leaf-HFs-A, leaf-HFs-B and leaf-HFs-C are 163, 119, and 251, respectively.
The proportions of leaf-HFs-A, leaf-HFs-B and leaf-HFs-C associated with both 21 $\mu$m and H$_{\alpha}$ structures are 58\% (95/163), 37\% (44/119) and 18\% (46/251), indicating that leaf-HFs-A are more evolved than leaf-HFs-C.
We also calculated the peak intensities of 7.7 $\mu$m, 21 $\mu$m and H$_{\alpha}$ emissions for each CO structure. As shown in Fig.~\ref{cp0} (f)-(h), all emissions gradually increase from leaf-HFs-C to leaf-HFs-A. 

\subsubsection{Scale and density}


\begin{figure}
\centering
\includegraphics[width=0.45\textwidth]{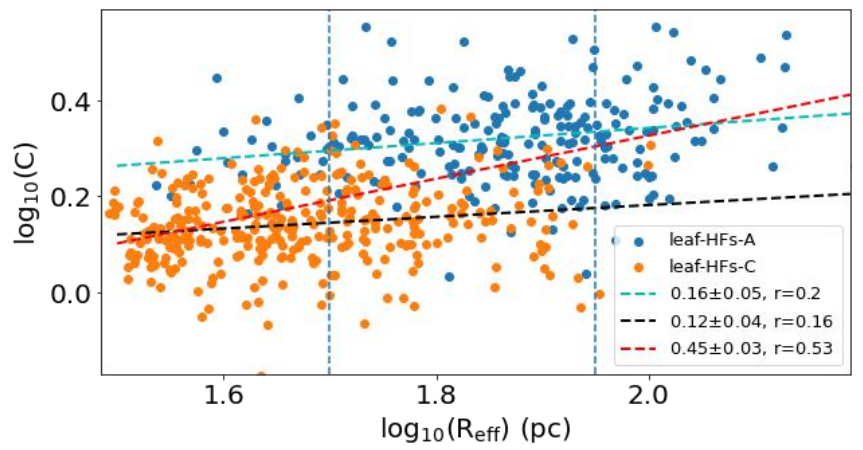}
\caption{Correlation between density contrast and scale (the effective radius) for leaf-HFs-A and leaf-HFs-C. The black, cyan, and red dashed lines fit leaf-HFs-C, leaf-HFs-A, and both leaf-HFs-C and leaf-HFs-A, respectively. ``r'' is the Pearson correlation coefficient.
Two vertical dashed lines mark the range where leaf-HFs-A and leaf-HFs-C have significant scale overlap.}
\label{smear}
\end{figure}

In Fig.~\ref{cp0}(a) and (e),
leaf-HFs-C have significantly smaller density contrasts and scales. However, leaf-HFs-C and leaf-HFs-B show comparable column density distributions in Fig.~\ref{cp0}(d), although leaf-HFs-B have larger density contrasts. Therefore,
Leaf-HFs-C are not necessarily low-density structures. It is just that the density distribution in leaf-HFs-C structures is relatively uniform. 
From these results,
one would argue that only when the structural scale is large enough, the density contrast can clearly manifest. In other words, the hub size of leaf-HFs-C could be too small to be resolved. Moreover, beam smearing effect may create an illusion of uniform density distribution. 
In order to check these possibilities, in Fig.~\ref{smear}, we fitted the correlation between density contrast and scale. Individually, for leaf-HFs-A and leaf-HFs-C, there is almost no correlation between density contrast and scale. When the two categories were combined, a correlation between density contrast and scale does appear. However, since leaf-HFs-A and leaf-HFs-C have significant scale overlap, if leaf-HFs-A can discern the hubs, leaf-HFs-C should be able to as well. Therefore, relative to leaf-HFs-A, the absence of the hubs in leaf-HFs-C is real and not merely a resolution issue. 

Two vertical dashed lines in Fig.~\ref{smear} mark the approximate range where leaf-HFs-A and leaf-HFs-C have significant scale overlap.
We confined the comparison of leaf-HFs-A, leaf-HFs-B and leaf-HFs-C to this scale range and obtained results consistent with previous findings, as shown in Fig.~\ref{cp1}. Therefore, the difference in scale is not a significant factor affecting the physical properties of different types of structures.


\subsection{Star formation}

\begin{figure}
\centering
\includegraphics[width=0.45\textwidth]{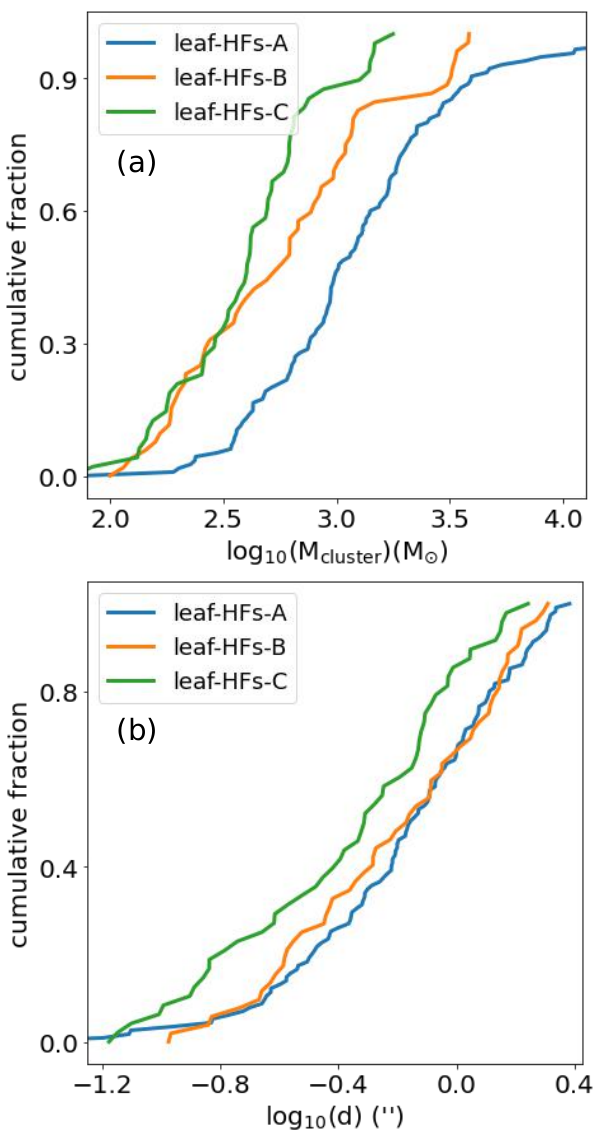}
\caption{
(a) The mass distribution of embedded star clusters;
(b) The spatial separations between the intensity peaks of CO and associated 21 $\mu$m structures.}
\label{mass}
\end{figure}

In this section, we examined the physical properties of embedded star clusters within molecular clouds.
Following the prescriptions described in \citet{Leroy2021-257}, the WISE 22 $\mu$m data can be used to calculate the local
star formation rate (SFR) surface density via
\begin{equation}
    \frac{\Sigma_{\rm SFR}}{\rm M_{\odot}~yr^{-1}~kpc^{-2}} = 3.8\times10^{-3} \left(\frac{I_{\rm 22 \mu m}}{\rm MJy~sr^{-1}}\right)\mathrm{cos}i,
    \label{}
\end{equation}
where $i$ is the inclination angle of the galaxy ($i$ = 0 degree is face-on).
In this work, we used JWST 21 $\mu$m data to estimate the SFR surface density. 
 
\citet{Kruijssen2014-439} and \citet{Kruijssen2018-479} have developed a statistically robust method to convert the observed spatial separations between cold gas and SFR tracers into their underlying timescales. This approach was employed by \citet{Kim2023-944} to study NGC 628 using emission maps of CO, JWST 21 $\mu$m, and H$_{\alpha}$ emission maps which respectively trace molecular clouds, embedded star formation, and exposed star formation. This analysis yielded systematic constraints on the duration of the embedded phase of star formation in NGC 628. They defined the duration of the embedded phase of star formation as the time during which CO
and 21 $\mu$m emissions are found to be overlapping, $t_{\mathrm{fb, 21 \mu m}}$. In contrast, the “heavily obscured phase” refers to the period when both CO and 21 $\mu$m emissions are present without associated H$_{\alpha}$ emission, $t_{\rm obsc}$. Finally, they obtained 
$t_{\mathrm{fb, 21 \mu m}} \approx 5.1^{+2.7}_{-1.4}$ Myr and $t_{\rm obsc} \approx 2.3^{+2.7}_{-1.4}$ Myr. In this work, we directly take the typical values $t_{\mathrm{fb, 21 \mu m}}=$ 5.1 Myr and $t_{\rm obsc}=$ 2.3 Myr. Therefore, the estimate is quite rough, given the significant uncertainty regarding the duration time.
For 185 CO structures associated with both 21 $\mu$m and H$_{\alpha}$ emissions, the duration time should be $\sim$2.3$-$5.1 Myr. The spatial separations $d$ between the intensity peaks of CO and 21 $\mu$m structures calculated in Sec.~\ref{association} can be used to distribute the age of the embedded star clusters in each molecular cloud. 
The typical value of $d$ is $\sim$ 0.64$''$, which is significantly larger than the typical positional uncertainty of each dataset.
Assuming $t_{0}$ = 2.3 Myr, the age of the embedded star cluster is
\begin{equation}
t = t_{0} + \frac{t_{\mathrm{fb, 21 \mu m}}-t_{\rm obsc}}{d_{\rm max}-d_{\rm min}} \times d,
\end{equation}
where $d_{\rm max}$ and $d_{\rm min}$ are the maximum and minimum spatial separations. By combining the local SFR of each molecular cloud, we estimated the total mass of the corresponding embedded star clusters. Fig.~\ref{mass}(a) shows the mass distribution of embedded star clusters. The cluster mass of leaf-HFs-A is much larger than those of leaf-HFs-B and leaf-HFs-C.
The results here are quite consistent with the findings in \citet{Hassani2023-944}.
For the JWST 21 $\mu$m compact source population identified in \citet{Hassani2023-944}, using spectral energy distribution (SED) fitting, they found that the 21 $\mu$m
sources have SEDs consistent with stellar population masses of 10$^{2}$ < $M_{*}/M_{\odot}$ < 10$^{4.5}$. They also found a range of mass-weighted
ages spanning 2–25 Myr, though most objects are < 8 Myr.

In Fig.~\ref{mass}(b), the spatial separations between the intensity peaks of CO and 21 $\mu$m structures of leaf-HFs-A are also larger than leaf-HFs-C. 
Since the spatial separation can measure the evolutionary timescales of molecular clouds, this is further evidence supporting that leaf-HFs-A is more evolved than leaf-HFs-C.
However, the sample size here is very limited. The numbers of leaf-HFs-A, leaf-HFs-B and leaf-HFs-C are 95, 44 and 46, respectively. More samples are necessary to quantitatively assess the correspondence between the separations and the evolutionary stages of molecular clouds.




\section{Discussion}

\subsection{Formation of hub-filament structures}

The hub-filament morphology could be shaped by either gravitational collapse \citep{Vazquez2019-490} or strong shocks due to turbulence \citep{Padoan2020-900}. 
In the second case, there is no anticipated consistency between the velocity gradients and the masses of the hubs, while in the first case, a relationship between the velocity gradients and the hub masses is naturally expected. Thus, the results presented in Sec.\ref{filament} clearly support the scenario in which the hub-filament structures form through gravitational contraction of gas structures on galaxy-cloud scales, which is also revealed in \citet{Zhou2024arXiv}. Similar findings on cloud-clump scales and clump-core scales are presented in \citet{Zhou2023-676} and \citet{Zhou2022-514}, also see \citet{Motte2018-56,Vazquez2019-490, Kumar2020-642} and references therein. 
The kinematic characteristics of gas structures on galaxy-cloud scale are very similar to those on cloud-clump and clump-core scales. 
Therefore, a picture emerges where dense cores act like hubs within clumps, clumps act like hubs within molecular clouds, and clouds act like hubs in spiral galaxies.
The interstellar medium from galaxy to dense core scales presents multi-scale/hierarchical hub-filament structures. Gas structures at different scales in the galaxy may be organized into hierarchical systems through gravitational coupling. 
The hierarchical hub-filament structures are also present in the galaxy cluster (hub)--cosmic web (filament) picture, reflecting the self-similarity of the structural organization in the universe. The fundamental reason may be that structures at different scales in the universe primarily evolve under the influence of gravity.

Hubs are essentially structures with higher density compared to the surrounding more diffuse gas. As local gravitational centers, these hubs seem to accrete the surrounding diffuse gas, forming hub-filament structures. The filamentary structures seem to be associated with gas flows converging towards the hubs. A hub-filament structure essentially consists of a gravitational center and the gas flow converging towards it. This could be considered as a fundamental structural type in the interstellar medium (ISM). In short, as long as there are local dense structures within the relatively diffuse ISM, hub-filament structures will inevitably form.

Finally, we note that
although the measured velocity gradients support the gravitational collapse of gas structures on galaxy-cloud scales, the collapse is much slower than a pure free-fall gravitational collapse. In the free-fall, the velocity gradient $\nabla v$ and the scale $R$ satisfy $\nabla v \propto R^{-1.5}$. However, Fig.\ref{gradient}(a) only gives $\nabla v \propto R^{-0.9}$. For NGC 4321 and NGC 5236, \citet{Zhou2024arXiv} obtained a similar slope, i.e.
$\nabla v \propto R^{-0.8}$. As discussed in \citet{Zhou2024arXiv}, the deviation from the
free-fall may come from the measurement biases, the coupling with the galactic potential, the tidal forces from the galaxy or the neighboring structures, or the turbulence and magnetic field supports \citep{Chevance2023-534}.

\subsection{Evolutionary sequence of molecular clouds}\label{}
\begin{figure*}
\centering
\includegraphics[width=1\textwidth]{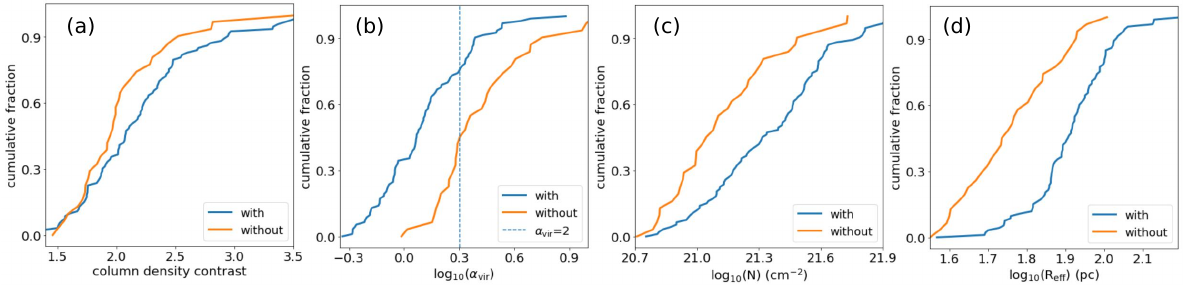}
\caption{Comparison of leaf-HFs-A with and without 21 $\mu$m and H$_{\alpha}$ emission. (a) Column density contrast; (b) Virial ratio. The vertical dashed line marks the position of $\alpha_{vir}=2$; (c) Column density; (d) Scale (the effective radius).}
\label{cpA}
\end{figure*}

\begin{figure}
\centering
\includegraphics[width=0.45\textwidth]{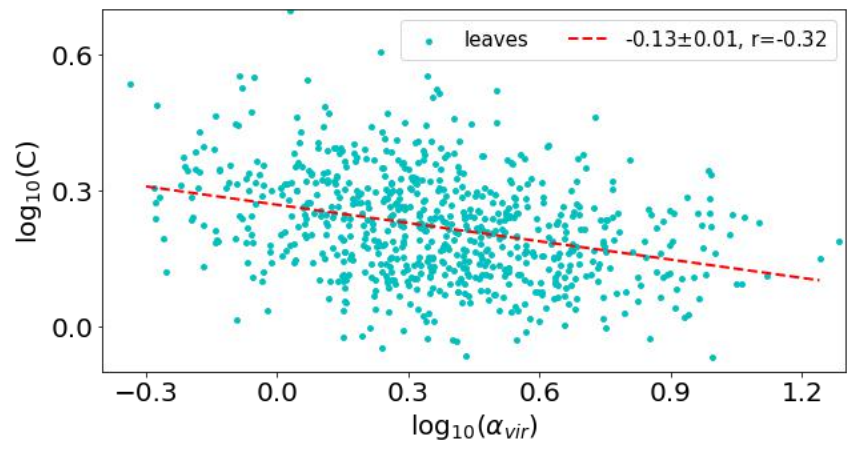}
\caption{Correlation between the density contrast and the virial ratio for all leaf structures. ``r'' is the Pearson correlation coefficient.}
\label{cv}
\end{figure}

The results of this work seem to suggest an evolutionary sequence among molecular structures. First, we focus solely on leaf-HFs-A. We compared two types of leaf-HFs-A, i.e. without and with 21 $\mu$m and H$_{\alpha}$ emissions. As shown in Fig.~\ref{cpA}, the structures with 21 $\mu$m and H$_{\alpha}$ emissions present larger density contrasts, lower virial ratios, higher column densities and larger scales (larger masses). In Fig.~\ref{cp0} and Fig.~\ref{cp1},
from leaf-HFs-C to leaf-HFs-A, we can see the same trend for each physical quantity.
Leaf-HFs-A without and with 21 $\mu$m and H$_{\alpha}$ emissions represent structures from two different evolutionary stages. 
Leaf-HFs-C and leaf-HFs-A may have the same relation.

A typical characteristic of hub-filament structures is that the hub region has significantly higher density compared to the surrounding filamentary structures, i.e. the density contrast. The hub-filament morphology is shaped by gravitational collapse.  Hub-filament structures only become apparent when gravitational collapse has progressed to a certain extent. In this work, the identified dense CO structures are potential hubs, which act as local gravitational centers. Leaf-HFs-A have significantly greater mass than C, so they serve as stronger gravitational centers. Leaf-HFs-A also present much better hub-filament morphology than leaf-HFs-C, which reflects the extent of gravitational collapse. Leaf-HFs-C have not yet exhibited a pronounced gravitational collapse process that would lead to a substantial density contrast. From leaf-HFs-C to leaf-HFs-A, gravitational collapse will increase the density and density contrast of the structure. Thus, the density contrast $C$ effectively gauges the degree of gravitational collapse and the depth of the gravitational potential well within a structure, which decisively shape the hub-filament morphology.

Fig.~\ref{cv} shows a clear correlation between the density contrast and the virial ratio. As expected, the density contrast decreases as the virial ratio increases. However, the classical virial analysis only considered the gravitational potential energy and internal kinetic energy may not completely reflect the true physical state of the identified structures. A more comprehensive virial analysis should include additional physical mechanisms, such as external pressure \citep{Spitzer1978,Elmegreen1989-338,Kirk2017-846,Li2020-896,Sun2020-892,Zhou2024-682}, tidal forces \citet{Ballesteros2009-393,Ballesteros2009-395,Ramirez2022-515,Li2024-528,Zhou2024-686} and magnetic fields \citep{Dib2007-661, Crutcher2010-725,Li2011-479,Crutcher2012-50,Seifried2020-497, Ibanez2022-925, Ganguly2023-525}. Local dense structures are embedded within larger gas environments. Their interactions with the surrounding environment may significantly impact their physical state. After calculating a more accurate virial ratio, the correlation in Fig.~\ref{cv} should be stronger.

As the masses and scales of the structures also increase from leaf-HFs-C to leaf-HFs-A, the results of this work suggest that the structures inevitably accrete matter from their surrounding environment during the evolution. 
It is consistent with the analysis in Sec.~\ref{filament} and the results presented in \citet{Zhou2024arXiv} for NGC 5236 (M83) and NGC 4321 (M100). In these spiral galaxies, there are clear velocity gradients around intensity peaks, and the variations in velocity gradient across different scales suggest a gradual and consistent increase in velocity gradient from large to small scales, indicative of gravitational collapse and gas inflow at different scales. Local dense structures act as local gravitational centers, naturally accreting matter from the surrounding diffuse gas environment and thereby accumulating mass.

\section{Summary}

We decomposed the spiral galaxy NGC 628 into multi-scale hub-filament structures using the CO (2$-$1) line map. The main conclusions are as follows:

1. The intensity peaks as potential hubs were identified  based on the integrated intensity (Moment 0) map of CO (2$-$1) emission by the dendrogram algorithm. Around all intensity peaks, we extracted the average spectra for all structures to decompose their velocity components and fix their velocity ranges. The final identification of hub-filament structures is based on the cleaned Moment 0 map of each structure made by restricting to the fixed velocity range to exclude the potential overlap of uncorrelated velocity components. In practice, some steps might not be necessary, as almost all structures show only one velocity component.

2. All leaf structures as potential hubs were classified into three categories, i.e. leaf-HFs-A, leaf-HFs-B and leaf-HFs-C. For leaf-HFs-C, the density contrast between the hub and the surrounding diffuse gas is not pronounced. Both leaf-HFs-A and leaf-HFs-B have clear central hubs. But leaf-HFs-A exhibit the best hub-filament morphology, which also have the highest density contrast, the largest mass and the lowest virial ratio. 

3. We employed the FILFINDER algorithm to identify and characterize filaments within 185 leaf-HFs-A structures using integrated intensity maps. We also fitted the velocity gradients around intensity peaks, a process performed after removing the global large-scale velocity gradients attributed to the galaxy's rotation. Measurements of velocity gradients provide evidence for gas inflow within these structures, which can also serve as kinematic evidence that these structures can be regarded as hub-filament structures.

4. Leaf-HFs-C are not necessarily low-density structures. It is just that their density distribution is relatively uniform. There may be an evolutionary sequence from leaf-HFs-C to leaf-HFs-A. Currently, leaf-HFs-C lack a distinct gravitational collapse process that would result in significant density contrast.
The numbers of the associated 21 $\mu$m and H$_{\alpha}$ structures and the peak intensities of 7.7 $\mu$m, 21 $\mu$m and H$_{\alpha}$ emissions decrease from leaf-HFs-A to leaf-HFs-C. The spatial separations between the intensity peaks of CO and 21 $\mu$m structures of leaf-HFs-A are larger than leaf-HFs-C. These evidence indicate that leaf-HFs-A are more evolved than leaf-HFs-C.

5. There is a clear correlation between the density contrast and the virial ratio, and the density contrast decreases as the virial ratio increases.
The density contrast $C$ effectively measures the extent of gravitational collapse and the depth of the gravitational potential well of the structure that shape the hub-filament morphology. In terms of reflecting the development and evolutionary stage of the structure, density contrast is more crucial than density itself.

6. Combining the local star formation rate (SFR) derived from the JWST 21 $\mu$m emission with the timescale revealed by the spatial separation between CO and 21 $\mu$m emissions yields mass estimates of embedded star clusters in molecular clouds comparable to those obtained from spectral energy distribution (SED) fitting. As expected, the cluster mass of leaf-HFs-A is much larger than those of leaf-HFs-B and leaf-HFs-C.

7. Combined with the kinematic evidence presented in \citet{Zhou2024arXiv}, a picture emerges where
molecular gas in spiral galaxies is organized into a network of structures through gravitational coupling of multi-scale hub-filament structures, consistent with the global hierarchical collapse scenario \citep{Vazquez2019-490}.
Molecular clouds with sizes of hundreds of pc in NGC 628 are knots in networks of hub-filament systems, and are the local gravitational centers and the main star-forming sites. 

\section*{Acknowledgements}
We would like to thank the referee for the detailed comments and suggestions that significantly improve and clarify this work.
It is a pleasure to thank the PHANGS team, the data cubes and other data products shared by the team make this work can be carried out easily. This paper makes use of the following ALMA data: 
ADS/JAO.ALMA\#2012.1.00650.S and ADS/JAO.ALMA\#2017.1.00886.L.
ALMA is a partnership of ESO (representing its member states), NSF (USA) and NINS (Japan), together with NRC (Canada), NSTC and ASIAA (Taiwan), and KASI (Republic of Korea), in cooperation with the Republic of Chile. The Joint ALMA Observatory is operated by ESO, AUI/NRAO and NAOJ. Based on observations taken as part of the PHANGS-MUSE large program (Emsellem et al. 2021). Based on data products created from observations collected at the European Organisation for Astronomical Research in the Southern Hemisphere under ESO programme(s) 1100.B-0651, 095.C-0473, and 094.C-0623 (PHANGS-MUSE; PI Schinnerer), as well as 094.B-0321 (MAGNUM; PI Marconi), 099.B-0242, 0100.B-0116, 098.B-0551 (MAD; PI Carollo) and 097.B-0640 (TIMER; PI Gadotti). This research has made use of the services of the ESO Science Archive Facility.
This work is based on observations made with the NASA/ESA/CSA James Webb Space Telescope. The data were obtained from the Mikulski Archive for Space Telescopes at the Space Telescope Science Institute, which is operated by the Association of Universities for Research in Astronomy, Inc., under NASA contract NAS 5-03127 for JWST. These
observations are associated with program 2107.

\section{Data availability}
All the data used in this work are available on the PHANGS team website.
\footnote{\url{https://sites.google.com/view/phangs/home}}.

\bibliography{ref}

\begin{thebibliography}{}
\expandafter\ifx\csname natexlab\endcsname\relax\def\natexlab#1{#1}\fi
\providecommand{\url}[1]{\href{#1}{#1}}
\providecommand{\dodoi}[1]{doi:~\href{http://doi.org/#1}{\nolinkurl{#1}}}
\providecommand{\doeprint}[1]{\href{http://ascl.net/#1}{\nolinkurl{http://ascl.net/#1}}}
\providecommand{\doarXiv}[1]{\href{https://arxiv.org/abs/#1}{\nolinkurl{https://arxiv.org/abs/#1}}}

\bibitem[{{Anand} {et~al.}(2021){Anand}, {Lee}, {Van Dyk}, {Leroy}, {Rosolowsky}, {Schinnerer}, {Larson}, {Kourkchi}, {Kreckel}, {Scheuermann}, {Rizzi}, {Thilker}, {Tully}, {Bigiel}, {Blanc}, {Boquien}, {Chandar}, {Dale}, {Emsellem}, {Deger}, {Glover}, {Grasha}, {Groves}, {S. Klessen}, {Kruijssen}, {Querejeta}, {S{\'a}nchez-Bl{\'a}zquez}, {Schruba}, {Turner}, {Ubeda}, {Williams}, \& {Whitmore}}]{Anand2021-501}
{Anand}, G.~S., {Lee}, J.~C., {Van Dyk}, S.~D., {et~al.} 2021, \mnras, 501, 3621, \dodoi{10.1093/mnras/staa3668}

\bibitem[{{Ballesteros-Paredes} {et~al.}(2009{\natexlab{a}}){Ballesteros-Paredes}, {G{\'o}mez}, {Loinard}, {Torres}, \& {Pichardo}}]{Ballesteros2009-395}
{Ballesteros-Paredes}, J., {G{\'o}mez}, G.~C., {Loinard}, L., {Torres}, R.~M., \& {Pichardo}, B. 2009{\natexlab{a}}, \mnras, 395, L81, \dodoi{10.1111/j.1745-3933.2009.00647.x}

\bibitem[{{Ballesteros-Paredes} {et~al.}(2009{\natexlab{b}}){Ballesteros-Paredes}, {G{\'o}mez}, {Pichardo}, \& {V{\'a}zquez-Semadeni}}]{Ballesteros2009-393}
{Ballesteros-Paredes}, J., {G{\'o}mez}, G.~C., {Pichardo}, B., \& {V{\'a}zquez-Semadeni}, E. 2009{\natexlab{b}}, \mnras, 393, 1563, \dodoi{10.1111/j.1365-2966.2008.14278.x}

\bibitem[{{Chevance} {et~al.}(2023){Chevance}, {Krumholz}, {McLeod}, {Ostriker}, {Rosolowsky}, \& {Sternberg}}]{Chevance2023-534}
{Chevance}, M., {Krumholz}, M.~R., {McLeod}, A.~F., {et~al.} 2023, in Astronomical Society of the Pacific Conference Series, Vol. 534, Protostars and Planets VII, ed. S.~{Inutsuka}, Y.~{Aikawa}, T.~{Muto}, K.~{Tomida}, \& M.~{Tamura}, 1, \dodoi{10.48550/arXiv.2203.09570}

\bibitem[{{Crutcher}(2012)}]{Crutcher2012-50}
{Crutcher}, R.~M. 2012, \araa, 50, 29, \dodoi{10.1146/annurev-astro-081811-125514}

\bibitem[{{Crutcher} {et~al.}(2010){Crutcher}, {Wandelt}, {Heiles}, {Falgarone}, \& {Troland}}]{Crutcher2010-725}
{Crutcher}, R.~M., {Wandelt}, B., {Heiles}, C., {Falgarone}, E., \& {Troland}, T.~H. 2010, \apj, 725, 466, \dodoi{10.1088/0004-637X/725/1/466}

\bibitem[{{Davis} {et~al.}(2013){Davis}, {Alatalo}, {Bureau}, {Cappellari}, {Scott}, {Young}, {Blitz}, {Crocker}, {Bayet}, {Bois}, {Bournaud}, {Davies}, {de Zeeuw}, {Duc}, {Emsellem}, {Khochfar}, {Krajnovi{\'c}}, {Kuntschner}, {Lablanche}, {McDermid}, {Morganti}, {Naab}, {Oosterloo}, {Sarzi}, {Serra}, \& {Weijmans}}]{Davis2013-429}
{Davis}, T.~A., {Alatalo}, K., {Bureau}, M., {et~al.} 2013, \mnras, 429, 534, \dodoi{10.1093/mnras/sts353}

\bibitem[{{Dewangan} {et~al.}(2020){Dewangan}, {Ojha}, {Sharma}, {Palacio}, {Bhadari}, \& {Das}}]{Dewangan2020}
{Dewangan}, L.~K., {Ojha}, D.~K., {Sharma}, S., {et~al.} 2020, \apj, 903, 13, \dodoi{10.3847/1538-4357/abb827}

\bibitem[{{Dib}(2023)}]{Dib2023-524}
{Dib}, S. 2023, \mnras, 524, 1625, \dodoi{10.1093/mnras/stad1904}

\bibitem[{{Dib} {et~al.}(2012){Dib}, {Helou}, {Moore}, {Urquhart}, \& {Dariush}}]{Dib2012-758}
{Dib}, S., {Helou}, G., {Moore}, T. J.~T., {Urquhart}, J.~S., \& {Dariush}, A. 2012, \apj, 758, 125, \dodoi{10.1088/0004-637X/758/2/125}

\bibitem[{{Dib} {et~al.}(2007){Dib}, {Kim}, {V{\'a}zquez-Semadeni}, {Burkert}, \& {Shadmehri}}]{Dib2007-661}
{Dib}, S., {Kim}, J., {V{\'a}zquez-Semadeni}, E., {Burkert}, A., \& {Shadmehri}, M. 2007, \apj, 661, 262, \dodoi{10.1086/513708}

\bibitem[{{Elmegreen}(1989)}]{Elmegreen1989-338}
{Elmegreen}, B.~G. 1989, \apj, 338, 178, \dodoi{10.1086/167192}

\bibitem[{{Emsellem} {et~al.}(2022){Emsellem}, {Schinnerer}, {Santoro}, {Belfiore}, {Pessa}, {McElroy}, {Blanc}, {Congiu}, {Groves}, {Ho}, {Kreckel}, {Razza}, {Sanchez-Blazquez}, {Egorov}, {Faesi}, {Klessen}, {Leroy}, {Meidt}, {Querejeta}, {Rosolowsky}, {Scheuermann}, {Anand}, {Barnes}, {Be{\v{s}}li{\'c}}, {Bigiel}, {Boquien}, {Cao}, {Chevance}, {Dale}, {Eibensteiner}, {Glover}, {Grasha}, {Henshaw}, {Hughes}, {Koch}, {Kruijssen}, {Lee}, {Liu}, {Pan}, {Pety}, {Saito}, {Sandstrom}, {Schruba}, {Sun}, {Thilker}, {Usero}, {Watkins}, \& {Williams}}]{Emsellem2022-659}
{Emsellem}, E., {Schinnerer}, E., {Santoro}, F., {et~al.} 2022, \aap, 659, A191, \dodoi{10.1051/0004-6361/202141727}

\bibitem[{{Ganguly} {et~al.}(2023){Ganguly}, {Walch}, {Seifried}, {Clarke}, \& {Weis}}]{Ganguly2023-525}
{Ganguly}, S., {Walch}, S., {Seifried}, D., {Clarke}, S.~D., \& {Weis}, M. 2023, \mnras, 525, 721, \dodoi{10.1093/mnras/stad2054}

\bibitem[{{Hassani} {et~al.}(2023){Hassani}, {Rosolowsky}, {Leroy}, {Boquien}, {Lee}, {Barnes}, {Belfiore}, {Bigiel}, {Cao}, {Chevance}, {Dale}, {Egorov}, {Emsellem}, {Faesi}, {Grasha}, {Kim}, {Klessen}, {Kreckel}, {Kruijssen}, {Larson}, {Meidt}, {Sandstrom}, {Schinnerer}, {Thilker}, {Watkins}, {Whitmore}, \& {Williams}}]{Hassani2023-944}
{Hassani}, H., {Rosolowsky}, E., {Leroy}, A.~K., {et~al.} 2023, \apjl, 944, L21, \dodoi{10.3847/2041-8213/aca8ab}

\bibitem[{{Henshaw} {et~al.}(2014){Henshaw}, {Caselli}, {Fontani}, {Jim{\'e}nez-Serra}, \& {Tan}}]{Henshaw2014}
{Henshaw}, J.~D., {Caselli}, P., {Fontani}, F., {Jim{\'e}nez-Serra}, I., \& {Tan}, J.~C. 2014, \mnras, 440, 2860, \dodoi{10.1093/mnras/stu446}

\bibitem[{{Henshaw} {et~al.}(2020){Henshaw}, {Kruijssen}, {Longmore}, {Riener}, {Leroy}, {Rosolowsky}, {Ginsburg}, {Battersby}, {Chevance}, {Meidt}, {Glover}, {Hughes}, {Kainulainen}, {Klessen}, {Schinnerer}, {Schruba}, {Beuther}, {Bigiel}, {Blanc}, {Emsellem}, {Henning}, {Herrera}, {Koch}, {Pety}, {Ragan}, \& {Sun}}]{Henshaw2020-4}
{Henshaw}, J.~D., {Kruijssen}, J.~M.~D., {Longmore}, S.~N., {et~al.} 2020, Nature Astronomy, 4, 1064, \dodoi{10.1038/s41550-020-1126-z}

\bibitem[{{Ib{\'a}{\~n}ez-Mej{\'\i}a} {et~al.}(2022){Ib{\'a}{\~n}ez-Mej{\'\i}a}, {Mac Low}, \& {Klessen}}]{Ibanez2022-925}
{Ib{\'a}{\~n}ez-Mej{\'\i}a}, J.~C., {Mac Low}, M.-M., \& {Klessen}, R.~S. 2022, \apj, 925, 196, \dodoi{10.3847/1538-4357/ac3b58}

\bibitem[{{Issac} {et~al.}(2019){Issac}, {Tej}, {Liu}, {Varricatt}, {Vig}, {Ishwara Chandra}, \& {Schultheis}}]{Issac2019}
{Issac}, N., {Tej}, A., {Liu}, T., {et~al.} 2019, \mnras, 485, 1775, \dodoi{10.1093/mnras/stz466}

\bibitem[{{Kim} {et~al.}(2023){Kim}, {Chevance}, {Kruijssen}, {Barnes}, {Bigiel}, {Blanc}, {Boquien}, {Cao}, {Congiu}, {Dale}, {Egorov}, {Faesi}, {Glover}, {Grasha}, {Groves}, {Hassani}, {Hughes}, {Klessen}, {Kreckel}, {Larson}, {Lee}, {Leroy}, {Liu}, {Longmore}, {Meidt}, {Pan}, {Pety}, {Querejeta}, {Rosolowsky}, {Saito}, {Sandstrom}, {Schinnerer}, {Smith}, {Usero}, {Watkins}, \& {Williams}}]{Kim2023-944}
{Kim}, J., {Chevance}, M., {Kruijssen}, J.~M.~D., {et~al.} 2023, \apjl, 944, L20, \dodoi{10.3847/2041-8213/aca90a}

\bibitem[{{Kirk} {et~al.}(2017){Kirk}, {Friesen}, {Pineda}, {Rosolowsky}, {Offner}, {Matzner}, {Myers}, {Di Francesco}, {Caselli}, {Alves}, {Chac{\'o}n-Tanarro}, {Chen}, {Chun-Yuan Chen}, {Keown}, {Punanova}, {Seo}, {Shirley}, {Ginsburg}, {Hall}, {Singh}, {Arce}, {Goodman}, {Martin}, \& {Redaelli}}]{Kirk2017-846}
{Kirk}, H., {Friesen}, R.~K., {Pineda}, J.~E., {et~al.} 2017, \apj, 846, 144, \dodoi{10.3847/1538-4357/aa8631}

\bibitem[{{Kruijssen} \& {Longmore}(2014)}]{Kruijssen2014-439}
{Kruijssen}, J.~M.~D., \& {Longmore}, S.~N. 2014, \mnras, 439, 3239, \dodoi{10.1093/mnras/stu098}

\bibitem[{{Kruijssen} {et~al.}(2018){Kruijssen}, {Schruba}, {Hygate}, {Hu}, {Haydon}, \& {Longmore}}]{Kruijssen2018-479}
{Kruijssen}, J.~M.~D., {Schruba}, A., {Hygate}, A. P.~S., {et~al.} 2018, \mnras, 479, 1866, \dodoi{10.1093/mnras/sty1128}

\bibitem[{{Kumar} {et~al.}(2020){Kumar}, {Palmeirim}, {Arzoumanian}, \& {Inutsuka}}]{Kumar2020-642}
{Kumar}, M.~S.~N., {Palmeirim}, P., {Arzoumanian}, D., \& {Inutsuka}, S.~I. 2020, \aap, 642, A87, \dodoi{10.1051/0004-6361/202038232}

\bibitem[{{Lang} {et~al.}(2020){Lang}, {Meidt}, {Rosolowsky}, {Nofech}, {Schinnerer}, {Leroy}, {Emsellem}, {Pessa}, {Glover}, {Groves}, {Hughes}, {Kruijssen}, {Querejeta}, {Schruba}, {Bigiel}, {Blanc}, {Chevance}, {Colombo}, {Faesi}, {Henshaw}, {Herrera}, {Liu}, {Pety}, {Puschnig}, {Saito}, {Sun}, \& {Usero}}]{Lang2020-897}
{Lang}, P., {Meidt}, S.~E., {Rosolowsky}, E., {et~al.} 2020, \apj, 897, 122, \dodoi{10.3847/1538-4357/ab9953}

\bibitem[{{Lee} {et~al.}(2023){Lee}, {Sandstrom}, {Leroy}, {Thilker}, {Schinnerer}, {Rosolowsky}, {Larson}, {Egorov}, {Williams}, {Schmidt}, {Emsellem}, {Anand}, {Barnes}, {Belfiore}, {Be{\v{s}}li{\'c}}, {Bigiel}, {Blanc}, {Bolatto}, {Boquien}, {den Brok}, {Cao}, {Chandar}, {Chastenet}, {Chevance}, {Chiang}, {Congiu}, {Dale}, {Deger}, {Eibensteiner}, {Faesi}, {Glover}, {Grasha}, {Groves}, {Hassani}, {Henny}, {Henshaw}, {Hoyer}, {Hughes}, {Jeffreson}, {Jim{\'e}nez-Donaire}, {Kim}, {Kim}, {Klessen}, {Koch}, {Kreckel}, {Kruijssen}, {Li}, {Liu}, {Lopez}, {Maschmann}, {Chen}, {Meidt}, {Murphy}, {Neumann}, {Neumayer}, {Pan}, {Pessa}, {Pety}, {Querejeta}, {Pinna}, {Rodr{\'\i}guez}, {Saito}, {S{\'a}nchez-Bl{\'a}zquez}, {Santoro}, {Sardone}, {Smith}, {Sormani}, {Scheuermann}, {Stuber}, {Sutter}, {Sun}, {Teng}, {Tre{\ss}}, {Usero}, {Watkins}, {Whitmore}, \& {Razza}}]{Lee2023-944}
{Lee}, J.~C., {Sandstrom}, K.~M., {Leroy}, A.~K., {et~al.} 2023, \apjl, 944, L17, \dodoi{10.3847/2041-8213/acaaae}

\bibitem[{{Leroy} {et~al.}(2021{\natexlab{a}}){Leroy}, {Schinnerer}, {Hughes}, {Rosolowsky}, {Pety}, {Schruba}, {Usero}, {Blanc}, {Chevance}, {Emsellem}, {Faesi}, {Herrera}, {Liu}, {Meidt}, {Querejeta}, {Saito}, {Sandstrom}, {Sun}, {Williams}, {Anand}, {Barnes}, {Behrens}, {Belfiore}, {Benincasa}, {Be{\v{s}}li{\'c}}, {Bigiel}, {Bolatto}, {den Brok}, {Cao}, {Chandar}, {Chastenet}, {Chiang}, {Congiu}, {Dale}, {Deger}, {Eibensteiner}, {Egorov}, {Garc{\'\i}a-Rodr{\'\i}guez}, {Glover}, {Grasha}, {Henshaw}, {Ho}, {Kepley}, {Kim}, {Klessen}, {Kreckel}, {Koch}, {Kruijssen}, {Larson}, {Lee}, {Lopez}, {Machado}, {Mayker}, {McElroy}, {Murphy}, {Ostriker}, {Pan}, {Pessa}, {Puschnig}, {Razza}, {S{\'a}nchez-Bl{\'a}zquez}, {Santoro}, {Sardone}, {Scheuermann}, {Sliwa}, {Sormani}, {Stuber}, {Thilker}, {Turner}, {Utomo}, {Watkins}, \& {Whitmore}}]{Leroy2021-257}
{Leroy}, A.~K., {Schinnerer}, E., {Hughes}, A., {et~al.} 2021{\natexlab{a}}, \apjs, 257, 43, \dodoi{10.3847/1538-4365/ac17f3}

\bibitem[{{Leroy} {et~al.}(2021{\natexlab{b}}){Leroy}, {Hughes}, {Liu}, {Pety}, {Rosolowsky}, {Saito}, {Schinnerer}, {Schruba}, {Usero}, {Faesi}, {Herrera}, {Chevance}, {Hygate}, {Kepley}, {Koch}, {Querejeta}, {Sliwa}, {Will}, {Wilson}, {Anand}, {Barnes}, {Belfiore}, {Be{\v{s}}li{\'c}}, {Bigiel}, {Blanc}, {Bolatto}, {Boquien}, {Cao}, {Chandar}, {Chastenet}, {Chiang}, {Congiu}, {Dale}, {Deger}, {den Brok}, {Eibensteiner}, {Emsellem}, {Garc{\'\i}a-Rodr{\'\i}guez}, {Glover}, {Grasha}, {Groves}, {Henshaw}, {Jim{\'e}nez Donaire}, {Kim}, {Klessen}, {Kreckel}, {Kruijssen}, {Larson}, {Lee}, {Mayker}, {McElroy}, {Meidt}, {Mok}, {Pan}, {Puschnig}, {Razza}, {S{\'a}nchez-Bl'azquez}, {Sandstrom}, {Santoro}, {Sardone}, {Scheuermann}, {Sun}, {Thilker}, {Turner}, {Ubeda}, {Utomo}, {Watkins}, \& {Williams}}]{Leroy2021-255}
{Leroy}, A.~K., {Hughes}, A., {Liu}, D., {et~al.} 2021{\natexlab{b}}, \apjs, 255, 19, \dodoi{10.3847/1538-4365/abec80}

\bibitem[{{Leroy} {et~al.}(2022){Leroy}, {Rosolowsky}, {Usero}, {Sandstrom}, {Schinnerer}, {Schruba}, {Bolatto}, {Sun}, {Barnes}, {Belfiore}, {Bigiel}, {den Brok}, {Cao}, {Chiang}, {Chevance}, {Dale}, {Eibensteiner}, {Faesi}, {Glover}, {Hughes}, {Jim{\'e}nez Donaire}, {Klessen}, {Koch}, {Kruijssen}, {Liu}, {Meidt}, {Pan}, {Pety}, {Puschnig}, {Querejeta}, {Saito}, {Sardone}, {Watkins}, {Weiss}, \& {Williams}}]{Leroy2022-927}
{Leroy}, A.~K., {Rosolowsky}, E., {Usero}, A., {et~al.} 2022, \apj, 927, 149, \dodoi{10.3847/1538-4357/ac3490}

\bibitem[{{Li}(2024)}]{Li2024-528}
{Li}, G.-X. 2024, \mnras, 528, L52, \dodoi{10.1093/mnrasl/slad149}

\bibitem[{{Li} \& {Henning}(2011)}]{Li2011-479}
{Li}, H.-B., \& {Henning}, T. 2011, \nat, 479, 499, \dodoi{10.1038/nature10551}

\bibitem[{{Li} {et~al.}(2020){Li}, {Zhang}, {Liu}, {Beuther}, {Palau}, {Girart}, {Smith}, {Hora}, {Lin}, {Qiu}, {Strom}, {Wang}, {Li}, \& {Yue}}]{Li2020-896}
{Li}, S., {Zhang}, Q., {Liu}, H.~B., {et~al.} 2020, \apj, 896, 110, \dodoi{10.3847/1538-4357/ab84f1}

\bibitem[{{Lindner} {et~al.}(2015){Lindner}, {Vera-Ciro}, {Murray}, {Stanimirovi{\'c}}, {Babler}, {Heiles}, {Hennebelle}, {Goss}, \& {Dickey}}]{Lindner2015-149}
{Lindner}, R.~R., {Vera-Ciro}, C., {Murray}, C.~E., {et~al.} 2015, \aj, 149, 138, \dodoi{10.1088/0004-6256/149/4/138}

\bibitem[{{Liu} {et~al.}(2022){Liu}, {Tej}, {Liu}, {Goldsmith}, {Stutz}, {Juvela}, {Qin}, {Xu}, {Bronfman}, {Evans}, {Saha}, {Issac}, {Tatematsu}, {Wang}, {Li}, {Zhang}, {Baug}, {Dewangan}, {Wu}, {Zhang}, {Lee}, {Liu}, {Zhou}, \& {Soam}}]{Liu2022-511}
{Liu}, H.-L., {Tej}, A., {Liu}, T., {et~al.} 2022, \mnras, 511, 4480, \dodoi{10.1093/mnras/stac378}

\bibitem[{{Liu} {et~al.}(2016){Liu}, {Zhang}, {Kim}, {Wu}, {Lee}, {Goldsmith}, {Li}, {Liu}, {Chen}, {Tatematsu}, {Wang}, {Lee}, {Qin}, {Mardones}, \& {Cho}}]{Liu2016}
{Liu}, T., {Zhang}, Q., {Kim}, K.-T., {et~al.} 2016, \apj, 824, 31, \dodoi{10.3847/0004-637X/824/1/31}

\bibitem[{{Liu} {et~al.}(2020){Liu}, {Evans}, {Kim}, {Goldsmith}, {Liu}, {Zhang}, {Tatematsu}, {Wang}, {Juvela}, {Bronfman}, {Cunningham}, {Garay}, {Hirota}, {Lee}, {Kang}, {Li}, {Li}, {Mardones}, {Qin}, {Ristorcelli}, {Tej}, {Toth}, {Wu}, {Wu}, {Yi}, {Yun}, {Liu}, {Peng}, {Li}, {Li}, {Lee}, {Shen}, {Baug}, {Wang}, {Zhang}, {Issac}, {Zhu}, {Luo}, {Soam}, {Liu}, {Xu}, {Wang}, {Zhang}, {Ren}, \& {Zhang}}]{Liu2020}
{Liu}, T., {Evans}, N.~J., {Kim}, K.-T., {et~al.} 2020, \mnras, 496, 2790, \dodoi{10.1093/mnras/staa1577}

\bibitem[{{Lu} {et~al.}(2018){Lu}, {Zhang}, {Liu}, {Sanhueza}, {Tatematsu}, {Feng}, {Smith}, {Myers}, {Sridharan}, \& {Gu}}]{Lu2018}
{Lu}, X., {Zhang}, Q., {Liu}, H.~B., {et~al.} 2018, \apj, 855, 9, \dodoi{10.3847/1538-4357/aaad11}

\bibitem[{{McKee} \& {Ostriker}(2007)}]{McKee2007-45}
{McKee}, C.~F., \& {Ostriker}, E.~C. 2007, \araa, 45, 565, \dodoi{10.1146/annurev.astro.45.051806.110602}

\bibitem[{{Motte} {et~al.}(2018){Motte}, {Bontemps}, \& {Louvet}}]{Motte2018-56}
{Motte}, F., {Bontemps}, S., \& {Louvet}, F. 2018, \araa, 56, 41, \dodoi{10.1146/annurev-astro-091916-055235}

\bibitem[{{Padoan} {et~al.}(2020){Padoan}, {Pan}, {Juvela}, {Haugb{\o}lle}, \& {Nordlund}}]{Padoan2020-900}
{Padoan}, P., {Pan}, L., {Juvela}, M., {Haugb{\o}lle}, T., \& {Nordlund}, {\r{A}}. 2020, \apj, 900, 82, \dodoi{10.3847/1538-4357/abaa47}

\bibitem[{{Peretto} {et~al.}(2013){Peretto}, {Fuller}, {Duarte-Cabral}, {Avison}, {Hennebelle}, {Pineda}, {Andr{\'e}}, {Bontemps}, {Motte}, {Schneider}, \& {Molinari}}]{Peretto2013}
{Peretto}, N., {Fuller}, G.~A., {Duarte-Cabral}, A., {et~al.} 2013, \aap, 555, A112, \dodoi{10.1051/0004-6361/201321318}

\bibitem[{{Ram{\'\i}rez-Galeano} {et~al.}(2022){Ram{\'\i}rez-Galeano}, {Ballesteros-Paredes}, {Smith}, {Camacho}, \& {Zamora-Avil{\'e}s}}]{Ramirez2022-515}
{Ram{\'\i}rez-Galeano}, L., {Ballesteros-Paredes}, J., {Smith}, R.~J., {Camacho}, V., \& {Zamora-Avil{\'e}s}, M. 2022, \mnras, 515, 2822, \dodoi{10.1093/mnras/stac1848}

\bibitem[{{Riener} {et~al.}(2019){Riener}, {Kainulainen}, {Henshaw}, {Orkisz}, {Murray}, \& {Beuther}}]{Riener2019-628}
{Riener}, M., {Kainulainen}, J., {Henshaw}, J.~D., {et~al.} 2019, \aap, 628, A78, \dodoi{10.1051/0004-6361/201935519}

\bibitem[{{Rosolowsky} {et~al.}(2008){Rosolowsky}, {Pineda}, {Kauffmann}, \& {Goodman}}]{Rosolowsky2008-679}
{Rosolowsky}, E.~W., {Pineda}, J.~E., {Kauffmann}, J., \& {Goodman}, A.~A. 2008, \apj, 679, 1338, \dodoi{10.1086/587685}

\bibitem[{{Seifried} {et~al.}(2020){Seifried}, {Walch}, {Weis}, {Reissl}, {Soler}, {Klessen}, \& {Joshi}}]{Seifried2020-497}
{Seifried}, D., {Walch}, S., {Weis}, M., {et~al.} 2020, \mnras, 497, 4196, \dodoi{10.1093/mnras/staa2231}

\bibitem[{{Spitzer}(1978)}]{Spitzer1978}
{Spitzer}, L. 1978, {Physical processes in the interstellar medium}, \dodoi{10.1002/9783527617722}

\bibitem[{{Sun} {et~al.}(2020){Sun}, {Leroy}, {Ostriker}, {Hughes}, {Rosolowsky}, {Schruba}, {Schinnerer}, {Blanc}, {Faesi}, {Kruijssen}, {Meidt}, {Utomo}, {Bigiel}, {Bolatto}, {Chevance}, {Chiang}, {Dale}, {Emsellem}, {Glover}, {Grasha}, {Henshaw}, {Herrera}, {Jimenez-Donaire}, {Lee}, {Pety}, {Querejeta}, {Saito}, {Sandstrom}, \& {Usero}}]{Sun2020-892}
{Sun}, J., {Leroy}, A.~K., {Ostriker}, E.~C., {et~al.} 2020, \apj, 892, 148, \dodoi{10.3847/1538-4357/ab781c}

\bibitem[{{V{\'a}zquez-Semadeni} {et~al.}(2019){V{\'a}zquez-Semadeni}, {Palau}, {Ballesteros-Paredes}, {G{\'o}mez}, \& {Zamora-Avil{\'e}s}}]{Vazquez2019-490}
{V{\'a}zquez-Semadeni}, E., {Palau}, A., {Ballesteros-Paredes}, J., {G{\'o}mez}, G.~C., \& {Zamora-Avil{\'e}s}, M. 2019, \mnras, 490, 3061, \dodoi{10.1093/mnras/stz2736}

\bibitem[{{Yuan} {et~al.}(2018){Yuan}, {Li}, {Wu}, {Ellingsen}, {Henkel}, {Wang}, {Liu}, {Liu}, {Zavagno}, {Ren}, \& {Huang}}]{Yuan2018}
{Yuan}, J., {Li}, J.-Z., {Wu}, Y., {et~al.} 2018, \apj, 852, 12, \dodoi{10.3847/1538-4357/aa9d40}

\bibitem[{{Zhang} {et~al.}(2015){Zhang}, {Wang}, {Lu}, \& {Jim{\'e}nez-Serra}}]{Zhang2015}
{Zhang}, Q., {Wang}, K., {Lu}, X., \& {Jim{\'e}nez-Serra}, I. 2015, \apj, 804, 141, \dodoi{10.1088/0004-637X/804/2/141}

\bibitem[{{Zhou} \& {Davis}(2024)}]{Zhou2024arXiv}
{Zhou}, J.~W., \& {Davis}, T.~A. 2024, arXiv e-prints, arXiv:2404.15862, \dodoi{10.48550/arXiv.2404.15862}

\bibitem[{{Zhou} {et~al.}(2024{\natexlab{a}}){Zhou}, {Dib}, {Juvela}, {Sanhueza}, {Wyrowski}, {Liu}, \& {Menten}}]{Zhou2024-686}
{Zhou}, J.~W., {Dib}, S., {Juvela}, M., {et~al.} 2024{\natexlab{a}}, \aap, 686, A146, \dodoi{10.1051/0004-6361/202449514}

\bibitem[{{Zhou} {et~al.}(2024{\natexlab{b}}){Zhou}, {Wyrowski}, {Neupane}, {Barlach Christensen}, {Menten}, {Li}, \& {Liu}}]{Zhou2024-682}
{Zhou}, J.~W., {Wyrowski}, F., {Neupane}, S., {et~al.} 2024{\natexlab{b}}, \aap, 682, A128, \dodoi{10.1051/0004-6361/202347377}

\bibitem[{{Zhou} {et~al.}(2022){Zhou}, {Liu}, {Evans}, {Garay}, {Goldsmith}, {G{\'o}mez}, {V{\'a}zquez-Semadeni}, {Liu}, {Stutz}, {Wang}, {Juvela}, {He}, {Li}, {Bronfman}, {Liu}, {Xu}, {Tej}, {Dewangan}, {Li}, {Zhang}, {Zhang}, {Ren}, {Tatematsu}, {Shing Li}, {Won Lee}, {Baug}, {Qin}, {Wu}, {Peng}, {Zhang}, {Liu}, {Luo}, {Ge}, {Saha}, {Chakali}, {Zhang}, {Kim}, {Ristorcelli}, {Shen}, \& {Li}}]{Zhou2022-514}
{Zhou}, J.-W., {Liu}, T., {Evans}, N.~J., {et~al.} 2022, \mnras, 514, 6038, \dodoi{10.1093/mnras/stac1735}

\bibitem[{{Zhou} {et~al.}(2023){Zhou}, {Wyrowski}, {Neupane}, {Urquhart}, {Evans}, {V{\'a}zquez-Semadeni}, {Menten}, {Gong}, \& {Liu}}]{Zhou2023-676}
{Zhou}, J.~W., {Wyrowski}, F., {Neupane}, S., {et~al.} 2023, \aap, 676, A69, \dodoi{10.1051/0004-6361/202346500}

\end{thebibliography}
\bibliographystyle{aasjournal}

\begin{appendix}
\twocolumn

\section{Supplementary maps}\label{map}

\begin{figure*}
\centering
\includegraphics[width=1\textwidth]{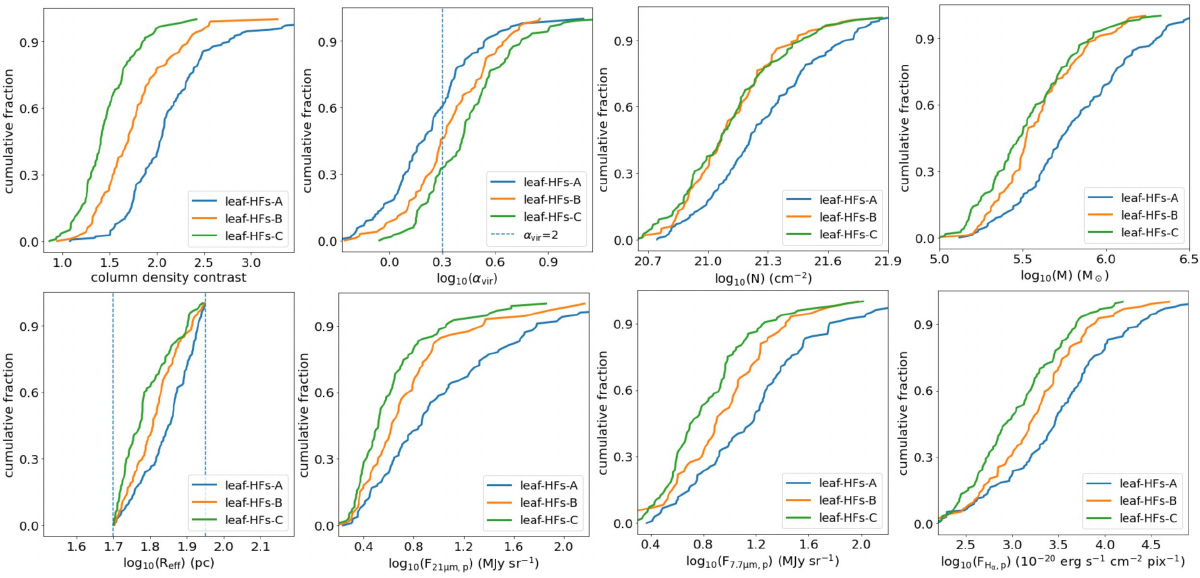}
\caption{Same as Fig.\ref{cp0}, but limited to the scale range marked by the vertical dashed lines in Fig.\ref{smear}.}
\label{cp1}
\end{figure*}

\end{appendix}

\clearpage
\noindent
\end{document}